\documentclass[a4paper,12pt,twoside,openright]{report}

\usepackage[american]{babel}
\usepackage{color,epsfig,subfigure,graphicx,amsmath,amssymb,bm}
\usepackage{fancyhdr}

\fancypagestyle{norm}{%      % Stile per i capitoli normali
\fancyhf{}  % azzera tutti i campi
\fancyfoot[C]{\thepage}
\fancyhead[LE]{\textit{\rightmark}}
\fancyhead[LO]{\textit{ \chaptername \ \thechapter}}
\fancyhead[RO]{\textit{\leftmark}}
\fancyhead[RE]{\textit{Section \ \thesection}}
}

\fancypagestyle{appendix}{%      % Stile per le appendici
\fancyhf{}  % azzera tutti i campi
\fancyfoot[C]{\thepage}
\fancyhead[LE]{\textit{\rightmark}}
\fancyhead[LO]{\textit{ Appendix \ \thechapter}}
\fancyhead[RO]{\textit{\leftmark}}
\fancyhead[RE]{\textit{Section \ \thesection}}
}

\fancypagestyle{appendixno}{%      % Stile per le appendici senza sezioni
\fancyhf{}  % azzera tutti i campi
\fancyfoot[C]{\thepage}
\fancyhead[LO,RE]{\textit{ Appendix \ \thechapter}}
\fancyhead[RO,LE]{\textit{\leftmark}}
}

\fancypagestyle{plain}{%         % Ridefinisce plain style per la prima pagina di un capitolo
\fancyhead{} % get rid of headers
}

\fancypagestyle{spec}{%     %Stile speciale per l'introduzione, toc...
\fancyhead{}
\fancyhead[LE]{\nouppercase{\textit{\rightmark}}}%nouppercase rende non maiuscolo i mark di bib e toc 
\fancyhead[RO]{\nouppercase{\textit{\rightmark}}}
\fancyfoot[C]{\thepage}
}

\newlength{\defbaselineskip}
\newcommand{\interlinea}[1]{\setlength{\baselineskip}{#1 \defbaselineskip}}

\newcommand\ket[1]{\left|#1\right\rangle}
\newcommand\bra[1]{\left\langle#1\right|}

\newcommand\be{\begin{equation}}
\newcommand\ee{\end{equation}}

\newcommand\bea{\begin{eqnarray}}
\newcommand\eea{\end{eqnarray}}

\begin{document}

%\maketitle

%%%% per una pagina nuova
\newpage
\thispagestyle{empty}
\cleardoublepage
%%%%%%%%%%%%%%%%%%%%%%%%%
%\frontmatter
%\input{frontespizio.tex}
\pagenumbering{roman}
\newpage
%%%%%%%%%%%%%%%%%%%%%%%%%%%% T A B L E  O F  C O N T E N T S  %%%%%%%%%%%%%%%%%%%%%%%%%%
\interlinea{1.3}
\pagestyle{spec}
\tableofcontents
%%%%%%%%%%%%%%%%%%%%%%%%%%%%%%%%%%%%%%%%%%%%%%%%%%%%%%%%%%%%%%%%%%%%%%%%%%%%%%%%%%%%%%%%

%%%% per una pagina vuota%
  \newpage				 	
  \thispagestyle{empty}	 	
  \cleardoublepage		 
%%%%%%%%%%%%%%%%%%%%%%%%%%

\pagenumbering{arabic}
%\mainmatter

% \typeout{Introduction}
% \chapter*{Introduction\markboth{Introduction}{Introduction}}
\chapter[Introducing Two-Dimensional Bose Gas]{Introducing Two-Dimensional Bose Gas}
% \addcontentsline{toc}{chapter}{Introducing Low Dimension Bose Gas}
\section{Low-dimensional Bose gases}

\label{Introduction}

\subsection{General overview}

The exotic phenomena of Bose-Einstein condensation was first predicted by 
Einstein \cite{einstein}, generalizing the concept of earlier work of Bose 
\cite{bose} on the quantum statistics of photons to indistinguishable 
particles with integral spin. These particles, later known to be bosons obey 
the symmetrical property of wave function under the exchange of particles. 
In nature atoms are classified into bosons or fermions according to their spin 
properties or intrinsic angular momentum. Bosons have integral spin while 
fermions are half-integral spin particles and are described by anti-symmetric 
wave function. Pauli's exclusion principles prohibiting two fermions to occupy 
a same quantum state is also due to this anti-symmetric property. On the other 
hand the symmetric nature of bosons allow them to macroscopically occupy a 
given state. 

The most illustrative example of bosonic behaviour of atoms is the process of 
Bose-Einstein Condensation (BEC) where at sufficiently low temperatures the 
bosons would become locked together in the lowest single-particle quantum 
state. Bosons that were distributed among many eigenstates plunge 
macroscopically into the equilibrium ground state (lowest energy state) 
via a phase transtition. At this stage the de Broglie waves of neighbouring 
atoms coalesce to form a giant matter-wave. For the three dimension (3D) 
geometry the condition for attaining BEC is to achieve a dimensionless phase 
space density of
\begin{equation}
\rho\lambda_{DB}^{3}\geq 2.612
\label{3Dcond}
\end{equation}
where $\tilde{\rho}$ and $\lambda_{DB}$ are the density of the system and 
de Broglie wave length respectively. One can also say that BEC occurs when 
the de Broglie wavelength becomes comparable with the interparticle 
distance $d=\rho^{-1/3}$.

BEC has been actively pursued over the last few decades in a variety of 
different systems. The initial drive in studying BEC focused on the 
intepretation of the collective (superfluid) behaviour of liquid $^4$He. In 
this case however, the interactions between atoms are so strong that the 
depletion of the condensate is very large. To investigate the physics of BEC 
phase transition in detail experimentalists have sought to obtain the 
phenomena of BEC in a weakly-interacting systems, where the interactions do 
not mask the quantum statistical effects.

\begin{figure}
\centering{
\epsfig{file=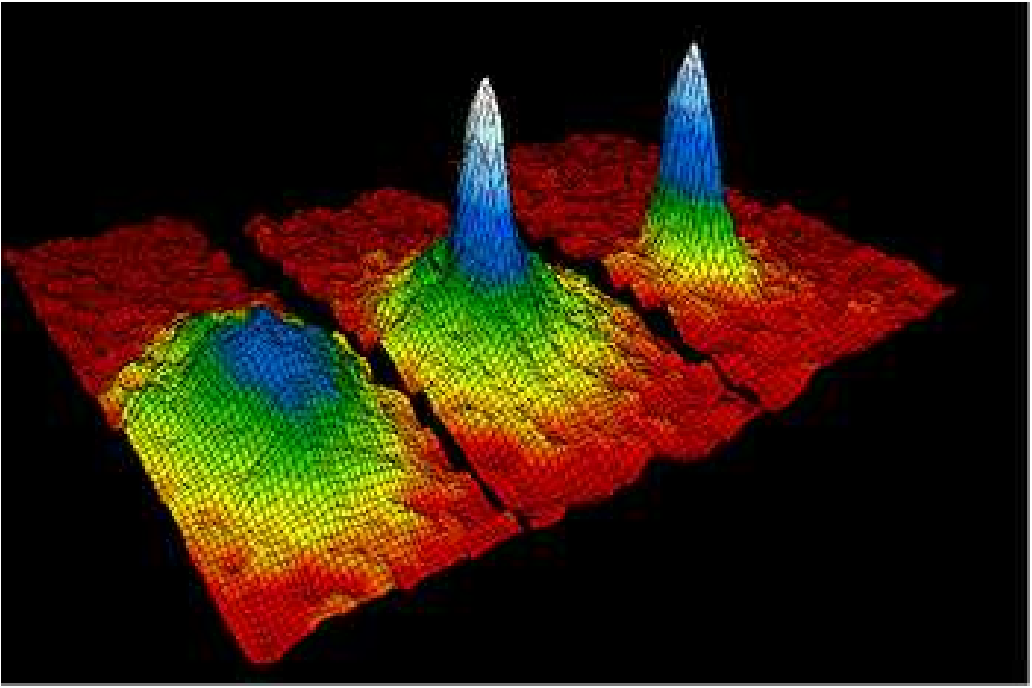,width=1.0\linewidth}}
\caption{The first BEC in a gas of $^{87}$Rb atoms, as it appears in 
two-dimensional time-of-flight images as temparature is decreased 
below $T_{c}$. Adapted from Anderson et. al. \cite{anderson} }
\label{fig_BEC}
\end{figure}

The first "true" Bose-Einstein condensate was created by Cornell, Wieman and 
co-workers at JILA on June 5, 1995 \cite{anderson}. They did this by cooling a 
dilute vapor consisting of approximately 2000 rubidium-87 atoms to 170 nK, the 
lowest temperature ever achieved at that time, using a combination of laser 
cooling (a technique that won its inventors Steven Chu, Claude 
Cohen-Tannoudji, and William D. Phillips the 1997 Nobel Prize in Physics) and 
magnetic evaporative cooling. The Laser cooling was followed by cycles of 
evaporative cooling, a principle similar to the thermalization of a cup of 
coffee. Henceafter a peak appears at the centre of the density distribution 
of the atomic cloud, which can be imaged by a resonant absorption technique
after a ballistic expansion of the cloud as it is realesed from the 
trap (See Fig. (\ref{fig_BEC})). 

About four months later, an independent effort led by Wolfgang Ketterle at MIT 
created a condensate made of sodium-23 \cite{davis}. Ketterle's condensate had 
about a hundred times more atoms, allowing him to obtain several important 
results such as the observation of quantum mechanical interference between two 
different condensates. Cornell, Wieman and Ketterle won the 2001 Nobel Prize 
for their achievement. In that same year Bose condensate on lithium was also 
observed by the group at Rice led by Randy Hulet \cite{bradley} 

The initial results by the JILA, MIT and Rice groups have led to an explosion 
of experimental activity. For instance, the first molecular Bose-Einstein 
condensates were created in 2003 by teams surrounding Rudolf Grimm \cite{rudy} 
at the University of Innsbruck and the first fermionic condensate by 
Deborah Jin and co-workers \cite{jin} at the University of Colorado 
at Boulder.

These pioneering realisation of BEC in a weakly-interacting system of 
trapped dilute alkali atoms in the mid 90's  has revived the interest in the 
theoretical studies of Bose gases. A rather massive amount of work has been 
done in the last couple of years, both to interpret the initial observations 
and to predict new phenomena. In the presence of harmonic confinement, the 
many-body theory of interacting Bose gases gives rise to several unexpected 
features. This opens new theoretical perspectives in this interdisciplinary 
field, where useful concepts coming from different areas of physics (atomic 
physics, quantum optics, statistical mechanics and condensed-matter physics) 
are merging together.

Due to the enormous breadth of the subject, it is quite impossible to give
a full coverage of the  basic theory of Bose-Einstein condensation and its 
experimental realizations together with its many novel applications. A large 
series of reviews and books has already emerged elaborating in clear 
details this novel state of matter. Readers are recommended to retrieve 
informations from Pethick and Smith \cite{pethick}, Legget \cite{legget}, 
Parkin and Walls \cite{parkin}, Griffin, Snoke and Stringari \cite{griffin} 
and so on. This thesis is written for a specialist audience for those whom 
already acquinted with the basic theory of Bose-Einstein condensation in a 3D 
geometry. In spite of this, or rather because of this I move ahead bypassing 
into the two-dimensional (2D) territories introducing the peculiarity 
phenomena arising in the squeezzed flat geometry.       

\subsection{Bose-Einstein condensate in the flat geometry}

The possibility of Bose-Einstein Condensation (BEC) phase transition occuring 
in a 2D geometry has a remarkable history and breakthroughs. The effect of 
dimensionality on the existence and character of BEC or superfluid phase 
transition has seen a spurt of interest in recent years. A peculiar feature of
the low-dimensionality (1D or 2D geometry) at finite temperature is the 
absence of a 'true condensate' following the Bogoliubov $k^{-2}$ theory, 
believed to be originated from the long wave-length phase fluctuations 
\cite{popov}. This has been proved in the seminal work of Mermin and 
Wagner \cite{mermin} and Hohenberg \cite{hohenberg67}.

Generally for a weakly interacting Bose gas at rest, the phenomenon of BEC is 
related to the macroscopic population of the single-particle state at zero 
momentum. One should expect a peak in the momentum distribution if 
condensation occur. One can also rely on the criterion first suggested by 
Onsager and Penrose \cite{penrose} in the context of quantum field theory. 
They associated BEC with the emergence of long-range ``off-diagonal'' order in 
the one-body density matrix 
$\rho({\bf r}-{\bf r'})=<\psi^{\dagger}({\bf r})\psi({\bf r'})>$. Here 
$\psi({\bf r})$ and $\psi^{\dagger}({\bf r})$ are the annihilation and 
creation field operators for a particle at position ${\bf r}$ and $<...>$ is 
the mean value. It was shown by Hohenberg \cite{hohenberg67} that for 
two-dimensional Bose gas the correlation function depends algebraically on 
$|{\bf r}-{\bf r'}|$ with an exponent $\gamma(T)$ that varies proportionally 
with the temperature $T$ as
\begin{equation}
<\psi^{\dagger}({\bf r})\psi({\bf r'})>\sim |{\bf r}-{\bf r'}|^{\gamma(T)}\,.
\label{aldecay}
\end{equation}

Algebraic decay of the correlation function is what one expect when the 
temperature is tuned to the critical temperature of a continuous phase 
transition. In fact, this is known as the Kosterlitz-Thouless transition where 
the system cools toward a superfluid state through the binding of vortices 
with opposite vorticity \cite{kosterlitz}. Nevertheless in the limit 
$T\rightarrow 0$ global coherence is achieved in a homogeneous 2D system and 
a true condensate then exists. In contrast, for a trapped 2D fluid the 
modification of the density of states caused by the confining potential 
allows a true condensate to exists even at finite temperature. 

Recent experiments have realized a quasi-two-dimensional (Q2D) cold Bose gas
by tuning the anisotropy of the trapping potential 
\cite{gorlitz,safanov,burger,schweik}. This achievement has attracted vast
attention on the new physics appearing in the currently studied
low- dimensional trapped gases and the nature of finite-size and 
finite-temperature effects.

\subsection{Homogeneous non-interacting Bose gas}

For a homogeneous gas, the individual particle eigenstates are characterized 
by momentum ${\bf k}$ and corresponding eigenenergy 
$\epsilon_{{\bf k}}=\hbar^2 {\bf k}^2/2m$, where $m$ is the mass of the 
particle. At thermal equilibrium with temperature $T$, a uniform gas of N 
bosons obeys the well-known Bose-Einstein distribution
\begin{equation}
f(\epsilon_{\bf k},T)=\frac{1}{\exp((\epsilon_{\bf k}-\mu)/k_{B}T)-1}\,,
\label{one-one}
\end{equation}
where $k_{B}$ is the Boltzmann constant and $\mu$ corresponds to the chemical 
potential of the system. The chemical potential can be interpreted as the 
energy required by a N-particle system to exchange one particle with its 
environment. The total number of particles in the excited states is given by 
the sum
\begin{equation}
N_{ex} = \sum_{\bf k} f(\epsilon_{\bf k},T) \,,
\label{one-two}
\end{equation}
which specifies the value of the chemical potential $\mu$. However, in a 
macroscopic system the volume is so large that the energy levels can be 
treated as continuous. Thus, Eq. (\ref{one-two}) can be written in the 
integral form 
\begin{equation}
N_{ex}= \tau \int d\epsilon_{\bf k}D(\epsilon_{\bf k})f(\epsilon_{\bf k},T)
\label{one-tree}
\end{equation}
with $\tau$ being the two-dimensional area and $D(\epsilon_{\bf k})$ is 
the density of state. In two-dimension, the smoothed density of state is 
$D(\epsilon_{\bf k})= L^2 m/2\pi\hbar^2$, where $L^2$ is the surface area. 
If we take the lower limit to be zero, then the integral does not exist
because of the divergence at low energies. In a two-dimensional box, the 
ground state has energy $\epsilon_{0}\sim \hbar^2/mL^2$, since its 
wavelength is comparable to the linear extent of the area. To estimate the 
number of excited particles, we implement a cut-off ($\epsilon_0$) in the 
integral as a lower bound. This gives us
\begin{equation}
N_{ex}\sim \frac{k_{B}T}{\epsilon_{0}}\ln (\frac{k_{B}T}{\epsilon_{0}})\,.
\label{one-four}
\end{equation}
The number of excited particles becomes equal to the total number of particles 
at a temperature (critical temperature) given by
\begin{equation}
k_{B}T_{c}\sim \frac{\hbar^2}{mL^2}\frac{N}{\ln N}=\hbar^2 
\frac{\tilde{\sigma}}{m\ln N}\,,
\label{one-five}
\end{equation}
where $\tilde{\sigma}=N/L^2$ is the number of particles per unit area. For a  
large system and in the thermodynamic limit ($N \rightarrow \infty$), the 
transition temperature tends to zero. This indicates that the macroscopic 
occupation of particles can only occur at zero temperature for a uniformly 
large system in the thermodynamic limit. Hohenberg \cite{hohenberg67} has 
proposed an $ad$ $absurdum$ reasoning for the absence of off-diagonal long 
range order for in a 2D uniform Bose gas at finite temperature. In the 
presence of condensate the following inequality
\begin{equation}
n({\bf k}) \geq -\frac{1}{2}+\frac{mk_{B}T}{\hbar^2 k^2}\frac{n_{0}}{n}
\label{one-six}
\end{equation}  
should hold, where $n({\bf k})$ is the momentum distribution at low momenta,
$n_{0}$ is the condensate density while $n$ is the total density. However 
for 2D, the density of excited particles evaluated using the integral 
$n_{exc}= \int d^2{\bf k}\, n({\bf k})$ diverges. So, one is led to the 
conclusion that the condensate fraction must be zero at any finite temperature.

\subsection{Kosterlitz-Thouless transition}

The theorem of Mermin and Wagner \cite{mermin}, which forbids the spontaneous 
breaking of a continuous symmetry in two-dimension due to the enhanced 
importance of long-wavelength thermal fluctuations, also reinforces the 
argument put forward by Hohenberg \cite{hohenberg67}. However, that a 
different kind of phase transition to a state with algebraic long-range order 
occurs in a 2D Bose fluid was first pointed out by Kosterlitz and 
Thouless \cite{kosterlitz}. This so called topological phase transition is 
associated with dissociation of a large number of vortex pairs and the 
immediate destruction of superfluidity caused by the phase slip process that 
occur once unbound vortices are present \cite{langer}. Nelson and 
Kosterlitz \cite{nelson77} predicted by means of a Renormalization-Group 
analysis that the discontinuous drop in the superfluid density $n_s$ at the 
critical temperature $T_c$ is of universal nature and such 
that $n_{s}\Lambda_{c}^2=4$, where $\Lambda_{c}$ is the thermal de Broglie 
wavelength.

The Kosterlitz-Thouless transition can easily be demonstrated using the 
two-dimensional XY-model. The model consists of planar rotors of unit length 
arranged on a two-dimensional square lattice. The Hamiltonian of the system is 
given by
\begin{equation}
H = -J \sum_{<i,j>} {\bf S}_i \,\cdot\,{\bf S}_j\,=-J\sum_{<i,j>}
\cos(\phi_i-\phi_j)\,
\label{one-seven}
\end{equation} 
where $J > 0$ and the sum $<i,j>$ is over the lattice sites of nearest 
neighbours only. Here we take $|{\bf S}_i|=1$  and $\phi_i$ is the angle that 
the $i$th spin makes with some arbitrary neighbours. Expanding about a local 
minima of $H$  gives
\begin{equation}
H-E_{0}\sim \frac{J}{2}\int d{\bf r} (\nabla\phi({\bf r}))^2
\label{one-eight}
\end{equation}
where $\phi({\bf r})$ is a phase function defined over the lattice sites and 
$E_{0}$ is the non-vortex ground state. By spherical symmetric consideration 
and taking into account a quantized vortex configuration, we can estimate the 
energy of a single vortex  $E_{vor}$ to be
\begin{equation}
E_{vor}-E_{0}\sim \pi J \ln (L/a_s) \,
\label{one-nine}
\end{equation}
where $L$ is the finite system size and $a_s$ is the lattice spacing. 
However, the free energy of a single vortex can be calculated using the 
Helmholtz free energy relation $F=E_{vor}-TS$, where $S$ is the entropy 
estimated from the number of places where we can position the vortex center
(in a classical viewpoint), namely on each of $L^2$ plaquette in a square 
lattice, $i.e.$ $S=k_{B}\ln(L^2/a^2)$. Accordingly the free energy is given by 
\begin{equation}
F= E_{0}+(\pi J-2k_{B}T)\ln (L/a_s)\,.
\label{one-ten}
\end{equation}
For $T<\pi\,J/2k_{B}$ the free energy will diverge to plus infinity as
$L\rightarrow \infty$. At temperatures $T>\pi\,J/2k_{B}$ the system can 
lower its free energy by producing vortices ($F\rightarrow -\infty$). 
This simple heuristic argument points to the fact that the logarithmic 
dependence on system size of the energy of the vortex combines with the 
logarithmic dependence of the entropy to produce the subtleties of 
the vortex unbinding transition. It is the logarithmic size dependence of the 
two-dimensional vortex energy that allows the outcome of the competition 
between the entropy and the energy to change qualitatively at a certain finite 
critical temperature $T_{KT}$ known as the Kosterlitz-Thouless transition.

In reality it is not the single vortices of the same signs that proliferate 
at a certain temperature. What happen is that the  larger vortex pairs which 
are bound together for temperature below $T_{KT}$ unbind at $T_{KT}$. This is
a collective effect.    
 
\subsection{Non-interacting trapped Bose gas}  

Let us consider a symmetric harmonic confining potential 
$V({\bf r})=m\omega^2(x^2+y^2)/2$ to describe the BEC cross-over of the 
2D Bose gas. In this case the particle energy is $E_{v}=\hbar\omega(n_x+n_y)$, 
with quantum numbers $n_x$ and $n_y$ which are non-negative integers. The 
density of states for this system is then $D(E)=E/(\hbar\omega)^2$, and the 
contribution of low-energy excited states to the total number of particles is 
negligible. The ground state ($E=0$) population is given by 
\begin{equation}
N_{0}= \frac{1}{\exp(-\mu/k_{B}T)-1}
\label{one-elv}
\end{equation}
which becomes macroscopic even for a small negative $\mu$. A cross-over to BEC 
regime exists in this case. Therefore, separating out the population of the 
ground state, one can replace the summation in Eq. (\ref{one-two}) by an 
integration, and one gets
\begin{equation}
N = N_{0}+ V \int d\epsilon_{\bf k}D(\epsilon_{\bf k})f(\epsilon_{\bf k},T)\,.
\label{one-tvelv}
\end{equation}
Using a the relation $-\mu/k_{B}T\approx 1/N_{0}$, which is obtained using 
Eq. (\ref{one-elv}) assuming that ground state is macroscopic populated
one can easily show that Eq (\ref{one-tvelv}) gives the following 
result \cite{petrov}:
\begin{equation}
N\left[ 1-(\frac{T}{T_c})^2 \right]=N_{0}-
(\frac{T}{\hbar\omega})^2 \frac{1+\ln N_{0}}{N_{0}}\,,
\label{one-tirteen}
\end{equation}
where $T_{c}=\sqrt{\frac{6N}{\pi^2}}\hbar\omega$ is the 2D BEC transition 
temperature for the non-interacting case. The effects of the interactions on 
this critical temperature will be discussed in chapter 2.

\section{Weakly interacting Bose gas}

Preliminary works of Bogoliubov \cite{bogoliubov}, Huang et al.
\cite{huang1,huang2} and Lee et al. \cite{Lee1,Lee2} were mainly focused on 
the theory of interacting homogeneous Bose gases. The extension to 
inhomogeneous gases was first explored by Gross \cite{gross61}, 
Pitaevskii \cite{pitaevskii1} and Fetter \cite{fetter71,fetter96}. 
The thermodynamic behaviour of trapped Bose gases has been already the object 
of several theoretical works, starting from the development of the 
Hartree-Fock formalism \cite{goldman}, the inclusion of interaction effects 
using the local density approximation \cite{oliva,chou} up to the most recent 
approaches based on self-consistent mean-field theory 
\cite{griffin,minguzzi97} and numerical simulation.  

\subsection{Weakly interacting regime}

I will consider weakly interacting gases with a short-range potential of 
interaction between particles. The total interaction energy is 
equal to the sum of pair interactions in the c-number formalism and can be 
written as $E_{int}= N^2g/2A$, where $g$ is the coupling constant for 
the pair interparticle interaction, N is the number of particles, and A is 
the area. Consequently, the interaction energy per particle is equal to 
$E_{int}=ng$, with $n$ being the $2D$ density. The general criteria for a 
weakly interacting regime assumes that the mean interparticle separation 
$\bar{r}$ greatly exceeds the characteristic radius of interaction between 
particles, $l_{e}$. In the 2D case we have $\bar{r}\sim (2\pi n)^{-1/2}$ 
and thus one obtain the inequality $n\,l_{e}^2\,\ll\,1$.

Another important condition for weakly interacting regimes is that the 
wavefunction is not influenced by the interaction between the particles if the 
interparticle distance is of order $\bar{r}$. Based on this condition, we 
can develop a physical picture which will be used for finding how the 
criterion of the weakly interacting regime depends on the dimensionality of 
the system. For example let consider a box of size $\bar{r}$, which 
accomodates one particle in average. In the limit $T\rightarrow 0$, the 
particle kinetic energy is $E_{kin}\sim \hbar^2/m\bar{r}^2$. The wavefunction 
is not influenced by the interparticle interaction if the kinetic energy per 
particle is much larger than the interaction energy per particle. The 
inequality $E_{kin}\gg\,E_{int}/N$ immeadiately gives the criterion of weakly 
interacting regime in term of of the density and coupling constant. In the 2D 
case this criterion takes the form
\begin{equation}
 \frac{m|g|}{2\pi\hbar^2}\ll \,1\,.
\label{one-forteen}
\end{equation} 
where $g$ is the 2D coupling constant which will be explained in detail in the 
next chapter. Most of the work in this thesis deals within the small 
repulsive/attractive limit ($g>0$ or $g<0$), except for the case where a 
weakly bound state of colliding atoms is present near resonance. Then the 
strength of $g$ is extremely large and even at very low densities the 
criterion of Eq. (\ref{one-forteen}) does not hold. Thus, one has to look for 
solutions beyond mean-field.

\subsection{Elementary excitations and quasiparticles}

The presence of a condensate mean field is naturally expected to affect the 
spectrum of the elementary excitations of a bosonic gas. This is a direct 
consequence of the fact that the atoms are not interacting with other 
individual atoms, but they are moving dominantly in the presence of a coherent 
condensate field. As expected, the deviation from the non-interacting 
spectrum become more pronounced as more particles are added to the 
condensate, since this lead to an increase in density. The notion of an 
excited (bare) atom interacting with condensed atoms leads quite 
generally to the idea of a dressed atom, or quasi-particle. One can see the 
significance of interactions to the change in the spectrum of elementary 
excitations by looking at the quasi-particle dispersion relation given by
\cite{bogoliubov}
\begin{equation}
\hbar\omega_{k}=\sqrt{\epsilon_{k}^2+2n_{c}V_{0}\epsilon_{k}}
\label{quasipen}
\end{equation}
where $\epsilon_{k}=\hbar^2k^2/2m$ corresponds to the non-interacting energy 
spectrum, $n_{c}$ is the condensate density and $V_{0}$ is the Fourier
transform of the interaction potential in the limit ${\bf k}\rightarrow 0$.

Recently, it was demonstrated by Al Khawaja  et. al \cite{khawaja} that by 
taking the following modifed type excitation spectrum: 
\begin{equation}
\hbar\omega_{k}=\sqrt{\epsilon_{k}^2+2\mu\epsilon_{k}}
\label{modqpen}
\end{equation}  
where $\mu=n_{c}T^{2B}(-2\mu)$, one is able to avoid the dangerous
ultraviolet and infrared divergences in the 2D Bose gas. $T^{2B}(-2\mu)$ 
is the multi-loop two-body T-matrix taken at energy argument $-2\mu$ which is 
exactly what it costs to excite two atoms out of the condensate 
(See Sec. 1.3 for more details). 

\subsection{Macroscopic framework of weakly interacting trapped Bose gas}

Let us consider the Bose gas trapped in a  2D isotropic planar trap 
$V({\bf r})=m\omega_{\perp}^2(x^2 + y^2)/2$. The starting point is the 
grand-canonical Hamiltonian in second quantization language:
\begin{equation}
H = H_{0}+\frac{1}{2}\int\,d{\bf r}\int\,d{\bf r'}\psi^\dagger({\bf r})
\psi^\dagger({\bf r'})V({\bf r}-{\bf r'})\psi({\bf r'})\psi({\bf r})
\label{one-sixteen}
\end{equation}    
where $H_{0}=\int\,d{\bf r} \psi^\dagger({\bf r})
\left[-\frac{\hbar^2}{2m}\nabla^2+V({\bf r})-\mu \right ]\psi({\bf r})$
, $\mu$ is the chemical potential and $V({\bf r}-{\bf r'})$ is the interatomic
interaction potential. The annihilation and creation operators are denoted by
$\psi^\dagger({\bf r})$ and $\psi({\bf r'})$ respectively and obey the Bose-
Einstein commutation relation
\begin{equation}
[\psi({\bf r}),\psi^{\dagger}({\bf r'})]=\delta({\bf r}-{\bf r'})\,\,\,\,\,
\,\,\,[\psi({\bf r}),\psi({\bf r'})]=[\psi^{\dagger}({\bf r}),
\psi^{\dagger}({\bf r'})]=0\,.
\label{tata}
\end{equation}
Since we are looking at system at very low temperature, only the contribution 
of s-wave scattering is important. Consequently we neglect the momentum 
dependence of the interatomic interaction and use 
$V({\bf r}-{\bf r'})=V_{0}\delta({\bf r}-{\bf r'})$. This leads to ultraviolet 
divergences but can be easily remedied. I will illustrate the technique in the 
next chapter. However with this contact potential the above Hamiltonian will 
reduce to the following one,
\begin{equation}
H = H_{0}+\frac{1}{2}\int\,d{\bf r}\,V_{0}\psi^\dagger({\bf r})
\psi^\dagger({\bf r})\psi({\bf r})\psi({\bf r})\,.
\label{one-7teen}
\end{equation}    
The time-dependent Heisenberg operator
 $\psi({\bf r},t)=\exp(iHt/\hbar)\psi({\bf r})\exp(-iHt/\hbar)$ obeys the 
equation of motion
\begin{equation}
i\hbar\frac{\partial \psi({\bf r},t)}{\partial t}=[\psi({\bf r},t),H]
\label{one-8teen}
\end{equation}
which yields the non-linear operator equation
\begin{equation}
i\hbar\frac{\partial \psi({\bf r},t)}{\partial t}= \left(-\frac{\hbar^2}
{2m}\nabla^2+V({\bf r}) \right )\psi({\bf r},t)+ g \psi^{\dagger}({\bf r},t)
\psi({\bf r},t)\psi({\bf r},t)\,.
\label{one-9teen}
\end{equation}
By separating out the condensate part of the particle field operator one can 
write $\psi({\bf r})$ as a sum of spatially varying condensate wavefunction
 $\Psi({\bf r})$ and a fluctuation field operator $\tilde{\psi}$,
\begin{equation}
 \psi= \Psi({\bf r})+ \tilde{\psi} \,.
\label{one-venti}
\end{equation}
The wave function $\Psi({\bf r})$ is also known as the order parameter and 
is defined as the statistical average of particle field operator
$\Psi({\bf r})=<\psi>$.

Using the decomposition in Eq.(\ref{one-venti}) above, one can write the 
interaction term $\psi^{\dagger}({\bf r},t)\psi({\bf r},t)\psi({\bf r},t)$ of
Eq.(\ref{one-9teen}) in the form
\begin{equation}
\psi^{\dagger}({\bf r},t)\psi({\bf r},t)\psi({\bf r'},t)=|\Psi|^2\Psi
+2|\Psi|^2\tilde{\psi}+\Psi^2\tilde{\psi}^{\dagger}+\Psi^{*}\tilde{\psi}
\tilde{\psi}+2\Psi\tilde{\psi}^{\dagger}\tilde{\psi}
+\tilde{\psi}^{\dagger}\tilde{\psi}\tilde{\psi} \,.
\label{one-venti2}
\end{equation}
Taking the average of the above relation one obtains
\begin{equation}
<\psi^{\dagger}({\bf r},t)\psi({\bf r},t)\psi({\bf r},t)>=n_{c}\Psi
+\tilde{m}\Psi^{*}+2n_{nc} 
+<\tilde{\psi}^{\dagger}\tilde{\psi}\tilde{\psi}> \,,
\label{one-venti2}
\end{equation}
where $n_{c}=|\Psi|^2$ is the condensate density, $n_{nc}=
<\tilde{\psi}^{\dagger}\tilde{\psi}>$ is the non-condensate density while
$\tilde{m}=<\tilde{\psi}\tilde{\psi}>$ is the off-diagonal (anomalous) density.
So taking the average for Eq. (\ref{one-9teen}) and using the relation in 
Eq. (\ref{one-venti2}) we arrive at the equation of motion for 
$\Psi({\bf r},t)$:
\begin{eqnarray}
i\hbar\frac{\partial \Psi({\bf r},t)}{\partial t}&=& \left(-\frac{\hbar^2}
{2m}\nabla^2+V({\bf r})+gn_{c}({\bf r},t)+2gn_{nc}({\bf r},t) \right )
\Psi({\bf r},t)\nonumber \\
&+& g\tilde{m}\Psi^{*}+g<\tilde{\psi}^{\dagger}\tilde{\psi}\tilde{\psi}>
\label{one-mama}
\end{eqnarray}

Neglecting the three-field correlation function $<\tilde{\psi}^{\dagger}\tilde{\psi}\tilde{\psi}>$ in the equation above is called the Hartree-Fock-Bogoliubov
(HFB) approximation. Further simplification to that by ignoring the term 
$\tilde{m}=<\tilde{\psi}\tilde{\psi}>$  is known as Hartree-Bogoliubov-Popov 
(HBF-P) approximation. In this thesis I will refer to the static Popov limit 
in which the fluctuation of the thermal cloud is ignored 
$n_{nc}({\bf r},t)\sim n_{nc}({\bf r})$ and will only consider a 
stationary solution for $\Psi({\bf r},t)=\Psi({\bf r})\exp(-i\mu\,t/\hbar)$. 
Thus the Non-linear Schr{\"o}dinger equation Eq. (\ref{one-mama}) leads to the
Gross-Pitaevskii (GP) equation 
\begin{equation}
 \left(-\frac{\hbar^2}{2m}\nabla^2+V({\bf r})+g_{2}n_{c}({\bf r})
+2g_{1} n_{nc}({\bf r}) \right )\Psi({\bf r})=\mu\Psi({\bf r})\,.
\label{one-GP}
\end{equation}
As anticipated, the general finite temperature GP equation above is not 
closed. Its solution requires one to know the information about the 
distribution of the non-condensate particles. Readers should also be aware 
that in the above equation I have introduced $g_{1}$ and $g_{2}$ which 
represent the coupling parameters due to condensate-noncondensate and 
condensate-condensate repulsion respectively. This choice of coupling 
parameters in studying the equilibrium property of condensate-thermal 
particles is the core issue of this thesis. Coming back to the problem of
thermal density distribution, I will take a semi-classical 
approximation \cite{giorgini} supplemented with a Hartree-Fock type energy 
spectrum
\begin{equation}
\epsilon({\bf p},{\bf r})=\frac{p^2}{2m}+V_{ext}({\bf r})+2g_{1}n_{c}
({\bf r})+2g_{1}n_{nc}({\bf r})\,.
\label{one-vent5}
\end{equation}
Hence, the thermal density is obtained by integrating out the momentum 
\begin{eqnarray}
n_{nc}({\bf r})&=&\int\frac{d{\bf p}}{(2\pi\hbar)^2}
f(\epsilon({\bf p},{\bf r}),T)
\nonumber \\
&=& -\frac{m\,k_{B}T}{2\pi\hbar^2}\times\ln\left[1-\exp\left(\beta(\mu-V_{eff}
({\bf r})\right)\right]\,.
\label{one-venty6}
\end{eqnarray}
where we consider the effective potential as $V_{eff}({\bf r})=V({\bf r})+
2g_{1}n_{c}({\bf r})+2g_{1}n_{nc}({\bf r})$. This type of method that couples 
solution of condensate and non-condensate density is also known as the 
Two-Fluid model.

\section{Brief introduction of the T-Matrix}

To describe the microscopic treatment of binary collisions, one generally 
considers the interaction between two condensed atom in a dilute 
weakly-interacting BEC. The complete set of two-particle interactions can be 
summed up into a single operator called the Transition matrix or simply 
T-matrix. This operator can be determined via a perturbation theory expansion 
in term of the actual interatomic potential. The possible scattering events 
can be described by a single interaction $\bra{{\bf k'}}V\ket{{\bf k}}$, or 
alternatively the particles may first make a transition to an intermediate 
state $\ket{{\bf q}}$ before interacting again to emerge in the state 
$\ket{{\bf k' }}$.  The repeated form of the scattering from initial to a 
final state via the intermediate state $\ket{{\bf q}}$ can be defined as an 
expansion of the T operators in term of the actual interatomic potential $V$ 
via
\begin{equation}
T =V+VG^{0}V+VG^{0}VG^{0}V+....
\label{defT}
\end{equation}
where $G^{0}=(z-H^{0})^{-1}$ corresponds to the unperturbed Green's function 
and $H^{0}$ is the free Hamiltonian in the absence of interactions while $z$ 
corresponds to the centre-of-mass-energy  of for the pair of atoms. In 
practice, the T operator can be assembled more explicitly in term of the 
Lippman Schwinger relation (see also in Sec. 2.1.1).
\begin{equation}
T =V+VG^{0}T \,.
\label{defT2}
\end{equation}
  
The two-body interaction above describe collision in $vacuo$ in which the 
intermediate states are single-particle in nature. In order to illustrate the 
interaction within a two-dimensional BEC we must consider the scattering in the
presence of condensate particles. Interactions of this nature can be described 
by the many-body T-matrix $T^{MB}({\bf k},{\bf k'},{\bf K};z)$ which due to 
the surrounding particles depends also on the center-of-mass momentum 
${\bf K}$ of the colliding particles. In the zero momentum limit we set 
${\bf k'},{\bf k},{\bf K}=0$. Since we are interested in the matrix element 
with all momenta equal to zero we find to a good approximation the following 
Bethe-Salpeter equation for the scattering of a quasi-particle,
\begin{eqnarray}
T^{MB}({\bf k},0,0;z)&=&V({\bf k})+\frac{1}{\tau}\sum_{{\bf k'}}V({\bf k'})
\frac{1+2N(\hbar\omega_{{\bf k'}})}{z-2\hbar\omega_{{\bf k'}}} \nonumber \\
&\times& T^{MB}({\bf k'},0,0;z)
\label{Bethe} 
\end{eqnarray}
where $\hbar\omega_{{\bf k}}$ is the quasi-particle energy 
defined in Eq. (\ref{modqpen}) and $N(x)=1/(\exp(\beta x)-1)$ is the 
Bose-Einstein distribution function. However in the zero momentum limit, 
the above Bethe-Salpeter equation is tainted by an ultraviolet divergence due 
to the neglect of the momentum dependence of the potential $V({\bf k})$. This 
divergence is remedied by using the following relation,
\begin{equation}
\frac{1}{T^{2B}(0,0;\bar{z})}=\frac{1}{V(0)}+\frac{1}{\tau}
\sum_{{\bf k}}\frac{1}{2\epsilon_{{\bf k}}-\bar{z}}
\label{remedy} 
\end{equation} 
where $\bar{z}=-j\mu$ with $j=1,2$ depending whether the collision is between 
two condensate atoms ($j=2$) or between one condensate and one non-condensate 
atom ($j=1$). The details are given in Sec. 2.1. The collision between 
condensate and the thermal atoms become more probable as temperature increases.
It is a crucial point and covers our investigation on the effect of 
interaction in calculating the density profiles at finite temperature 
(See Chapter 2). 

\subsection{Quantum control of the interacting Bose gas}

\begin{figure}
\centering{
\epsfig{file=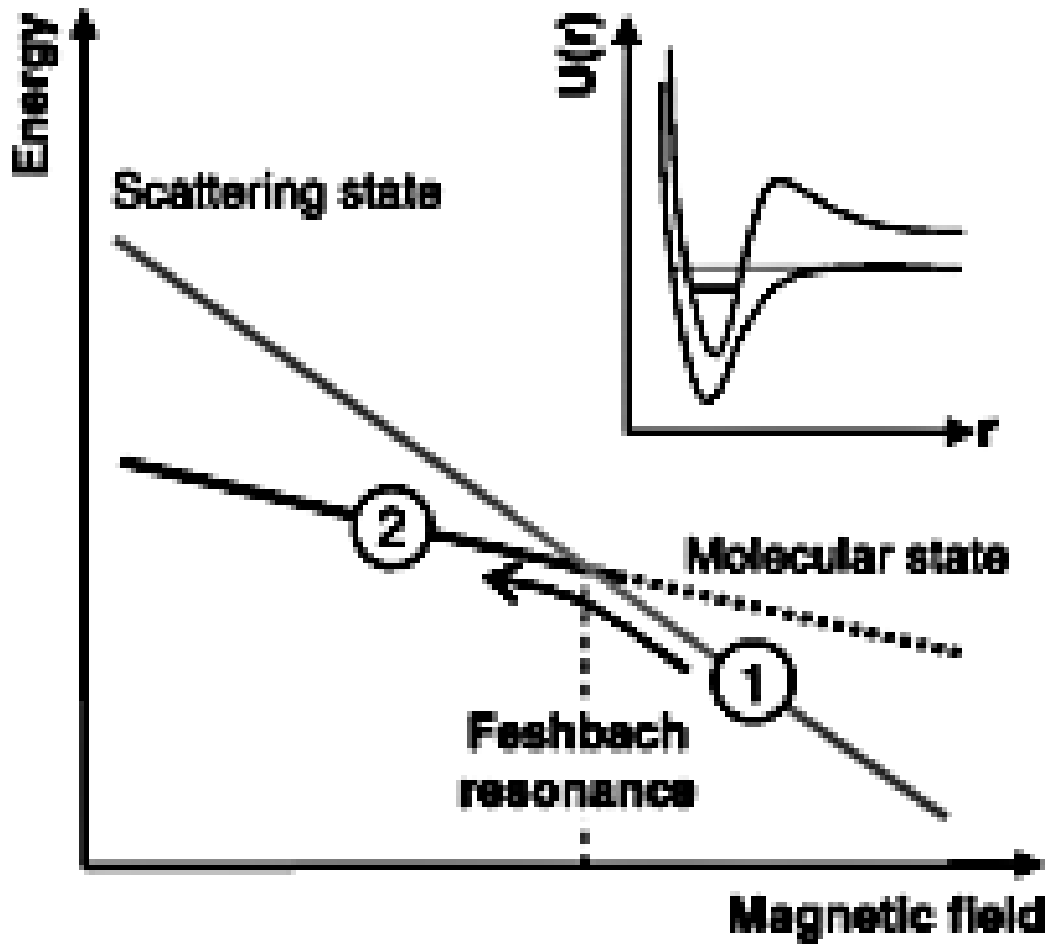,width=1.0\linewidth}}
\caption{Energy diagram for the atomic scattering state and the molecular bound
state. The Feshbach resonance condition occurs near 20 G, where the Zeeman 
energy of the atomic scattering state becomes equal to that of a molecular 
bound state because of the difference in magnetic moments. Molecules at (2) 
are created from the BEC at (1) by a downward sweep of the magnetic field 
across the resonance. For detection, a reversed sweep brings the molecules 
above the dissociation limit. The inset schematicallyshows the molecular 
potential that corresponds to the open channel (lower curve) and the molecular 
potential that supports the bound state (upper curve). U, potential energy; 
r, interatomic distance. From Herbig et. al. \cite{rudy}}
\label{fig_molecule}
\end{figure}

A Feshbach resonance occurs when the energy of a metastable molecular state 
$E_{res}$  matches the energy of a pair of incoming condensate atoms 
$E_{th}$ and a coupling exists between them during a collisional process 
\cite{moerdijk95}. As described in the previous section, the scattering 
of particles with zero relative energy or momentum is specified by the 
scattering amplitude related to the T matrix. If the energy of the 
incoming particles $E_{th}$ is close to the resonance energy $E_{res}$ of one 
particular bound state (resonance state $\ket{\psi_{res}}$), the contributions
from all other states will vary slowly with energy and they may be represented 
by non-resonant scattering length $a_{bg}$ whose energy dependence is much 
weaker compared to the resonance term. By writing the T matrix elements as a 
function of scattering length and using the argument discussed above we obtain 
the following relation,
\begin{equation}
  T^{MB}(0,0,0) = T^{MB}_{bg}(0,0,0) + \frac{C(B)}{E_{th}-E_{res}} \,.
\label{arelation}
\end{equation}
where $T^{MB}_{bg}$ is the off-resonant scattering amplitude and $C(B)$ is the 
resonant width, weakly dependent on $B$. The total scattering amplitude 
that includes resonant and non-resonant terms for a 3D Bose can be easily 
calculated as $T^{MB}(0,0,0)\sim T^{2B}(0,0,0)= 4\pi\hbar^2 a/m $ by the 
two-body scattering of particles within the Born approximation \cite{pethick}.
Eq. (\ref{arelation}) displays the energy dependence of the two-body T matrix 
and is characteristic of a Feshbach resonance. Hence the atomic interactions 
may be tuned by exploiting the fact that energies of the states depend on 
external parameters among which are the strength of a magnetic field. 
Expanding the energy denominator about the value of magnetic field we find
\begin{equation}
E_{th}-E_{res}\approx (\mu_{res}-2\mu_{i})(B-B_{o}),
\label{8teen}
\end{equation}
where $\mu_{i}=-\frac{\partial\epsilon_{i}}{\partial B}\,,i=1,2$
are the magnetic moments of the two atoms in the open channel while
$\mu_{res}=-\frac{\partial E_{res}}{\partial B}$ is the magnetic
moment of the molecular bound state. Substituting Eq. (\ref{8teen}) into 
Eq. (\ref{arelation}) one obtains \cite{moerdijk95}
\begin{equation}
a(B)= a_{bg}\left( 1-\frac{C(B)}{B-B_{0}} \right)\,.
\label{scatt}
\end{equation}
where $a_{bg}$ is the off-resonant scattering length. By sweeping across the 
magnetic field around the resonance location $B_{0}$ the coupling strength of 
the condensate can be used as a controlling tool. Hence one has the 
possibility not only to control the scattering length from a strongly 
repulsive to a strongly attractive regimes but also in obtaining a novel 
molecular condensate \cite{rudy} (as depicted in Fig (\ref{fig_molecule}).

Whether we can have a similar control of the interaction parameters for the 
strictly 2D condensate when the {\it{s}}-wave scattering length becomes larger 
than the width of the axial trapping \cite{schick,popov} will be addressed in 
the second part of chapter 2. To my knowledge, I have not encountered any 
attempts to study the Feshbach resoances for a strictly 2D Bose gas. This 
motivated me to look into such phenomena in those low-dimensional regime by 
using the similar technique conventionally used in studying them in the 3D 
case.

Once a tunable coupling parameter is calculated it will be incorporated into
the 2D dynamical GP equation to study the effect of the interactions on the 
density profiles at absolute zero temperature. I will perform a similar 
analysis as done by Kagan $et.al$ \cite{kagan2} by modifying the coupling 
strength as a dimensionless physical parameter $\eta$ which is defined as the 
ratio between the mean-field interaction energy and the harmonic-oscillator 
level spacing. The interaction is of repulsive nature if $\eta >0$, else it is 
attractive for $\eta <0$. The magnitude $\eta\ll1$ indicates that we are in 
the weakly repulsive (attractive) regime and $\eta\gg 1$ means a strongly
repulsive (attractive) for $\eta >0$ ($\eta<0$).

\subsection{The attractive interaction limit}

For an attractive interaction between particles ( $\eta<0$ ), the picture 
changes drastically. A Bose condensate with $|\eta|\gg1$, in which 
the structure of the trap levels is not important, cannot be formed at all, 
since in this case the accumulation of particles in one quantum state would be 
associated with increase in energy \cite{kagan2}. In this case the dominant 
elastic pair collision will prevent the formation of a Bose condensate with 
densities $(n_{c}/n_{o})\eta\gg 1)$, since the structure of trap levels is 
essentially smeared out by interatomic interaction and there is no gap for the 
excitation of particles from the condensate.

As a simple illustration, by taking the TF approximation to Eq. (\ref{one-GP})
(assuming $n_{nc}=0$ at $T=0$ ) in this limit ($|\eta| \gg 1$) and 
substituting $\eta = -k^2$ where $k$ is any positive number, we obtain the 
relation  
\begin{equation}
|\frac{\Psi({\bf r})}{\Psi(0)}|\sim \sqrt{1+\frac{1}{2}(\frac{r}{k})^2}\,.
\label{wentynove}
\end{equation}

What we have obtained is a monotonically increasing function which is not
physical. The use of Eq.(\ref{one-GP}) is inappropriate to analyse the 
behaviour of a condensate with attractive interaction. 
Consequently, we need to incorporate the physics of atom losses due to 
three-body recombination and formation of metastability condensate in the
attractive regime as investigated by several authors \cite{kagan2,saito,sadhan,savage} for the 3D case.\\ 
In order to model atom loss due to three-body recombination we add a 
phenomenological term proportional to the density squared $|\Psi(x)|^2$ with
coefficient $K_{3}/2$ \cite{huse}. The GP equation which incorporates a 
three-body loss term reads  
\begin{equation}
 i\hbar\frac{\partial \Psi(r,t)}{\partial t}=\left[-\frac{\hbar^2}{2m}
\nabla^2 + V_{ext}+g n_{c}-\frac{i\hbar K_{3}}{2}n_{c}^2\right] \Psi(r,t)\,,
\label{trenta}
\end{equation}
where the rate coefficient $K_{3}$ is related to the atom loss rate given by :
\begin{equation}
\frac{dN}{dt} = - K_{3} \int n_{c}^3 d{\bf r} \,.
\label{trentuno}
\end{equation}

A meta-stable solution of the above Eq. (\ref{trenta}) exist only if one 
chooses a number $N$ such that $N<N_{c}$ where $N_{c}$ is the critical number 
as reported for the 3D case \cite{baym96,ueda1,kim,antonius}. Beyond this 
number, the system would begins to collapse and lead to infinite density 
fluctuation to form hydrodynamic 'Black hole' \cite{antonius}. The 
calculation of this critical number for the 2D case is reported in 
chapter 2.   

\section{Dynamics of vortices in a 2D Bose fluid}
\label{many vortex}

Under appropriate stabilization conditions, such as steady applied rotation, 
vortices can form a regular array. In a rotating uniform superfluid, quantized 
vortex lines parallel to the axis of rotation form a Wigner-type crystal 
lattice \cite{andro}. At non-zero temperature, dissipative 
mutual friction from the normal component ensures that the array rotates with 
a  same angular velocity as the container. A triangular vortex array is 
energetically favoured for vortices near the rotation axis of rapidly 
rotating vessels of superfluid helium, (see Tkachenko \cite{tkac1,tkac2}). 

However, the properties of vortices in a Bose-Einstein atomic condensate atoms 
have proven to be more tractable compared to the superfluid helium. Following 
the creation of single vortex lines in an atomic condensate by a phase 
imprinting technique \cite{matthews}, large arrays have been created via 
mechanical rotation and stirring of the condensate with laser 
beam \cite{madison1,madison2,abo}. The recent observation of Tkachenko modes 
(collective excitations) in the Bose-condensed $^{87}Rb$ condensate by 
Coddington et al. \cite{coddington} stimulated more work on this direction. 
Congruent to that, there has been an enormous number of theoretical studies in 
analysing the properties of vortex lattices \cite{ho,wilkin,cooper,fischer1}, 
Tkachenco lattice oscillations (waves)\cite{baym2003,cozzini}, the transition 
into a novel properties of Fractional Quantum Hall Effects (FQHE) and 
strongly correlated uncondensed liquid \cite{fischer2,sinova}.

Atomic Bose condensates allow one to study superfluids over a range of 
rotational frequencies $\Omega$ \cite{madison2,abo,haljan}. Baym 
\cite{baym2003} has classified four distinct physical regimes as $\Omega$ 
increases: (1) The 'stiff' Thomas-Fermi regime, where $\Omega$ is 
small compared with the lowest compressional frequencies, $sk_{0}$, where 
$s$ is the sound velocity, $k_{0}\sim 1/R$ is the lowest wave number in the 
finite geometry, and $R$ is the size of the system transverse to the rotation 
axis. The system is an incompressible fluid in this regime responding 
effectively to rotation. (ii) In the limit $sk_{0}\,\ll\Omega\ll\,ms^2$, 
the system is said to be in a 'soft' Thomas-Fermi regime. Here the vortex core 
size is much smaller than inter-vortex spacing. (iii) When $\Omega\gg\,ms^2$, 
we enter a fast rotation regime where one is ensured that for a large vortex 
density, the intervortex separation is greater than the coherence length. This 
idea follows from the work of Baym \cite{baymcon} who showed shrinking of 
vortex cores when such limit is reached. In this rapidly rotating regime the 
system enters a effectively two-dimensional strong field. Viefers et al. 
\cite{viefers} have shown that as the angular momentum of N particle system 
gets larger, particles are spread out in space and the 'Yrast line' 
(the lowest possible interaction energy for a given angular momentum L) 
decreases. As the Yrast line decreases, the density of particle 
also decreases in parallel. Ho \cite{ho} on the other hand has predicted that 
in this limit particles should condense into the Lowest Landau Level (LLL) in 
the Coriolis force, similarly to charged particles in the quantum Hall regime. 
At relatively low vortex density, the vortices form a triangular 
lattice \cite{tkac1,tkac2} which is energetically favourable. There is a gap 
in the energy levels $\sim 2\omega_{\perp}$ when LLL acheived. The single 
particle wave-function in the LLL has the form $\phi_{i}(r)\sim z^{i}
\exp(-r^2/2d_{\perp}^2)$ where $z=x+iy$, i=0,1,2... and $d_{\perp}$ is the 
transverse oscillator length given by $\sqrt{\hbar/m\omega_{\perp}}$. The LLL 
condensate wave-function is a linear superposition of such states :
\begin{equation}
\Phi_{LLL}(r)\sim\sum_{i}c_{i}\,z^{i}\exp(- r^2/2d_{\perp}^2)
\sim\prod_{i=1}^{N}(z-z_{i})\exp(- r^2/2d_{\perp}^2)
\label{tre-rat}
\end{equation}
(iv) Eventually the vortex lattice melts \cite{rozhkov,sinova} and the 
systems enters a strongly correlated vortex liquid 
\cite{wilkin,viefers,regnault,fischer2}. The melting is somehow controlled by 
a dimensionless Lindemann parameter
\begin{equation}
c_{L}=\frac{\Omega}{\pi \rho}=\left(\frac{d}{l_{v}}\right)^2=\frac{n_{v}}
{\rho}
\label{tre-chimp}
\end{equation}
where $d\equiv \rho^{-1/2}$ is the inter-particle distance,
$l_{v}=\sqrt{\pi/\Omega}$ is the mean inter-vortex spacing 
and $n_{v}$ ( $n_{v}=\Omega/\pi$ ) is the density of vortices. 

This parameter controls the relative strength of the quantum fluctuations and 
the melting transition into quantum Hall regime at zero temperature. The 
melting filling factor calculated using various methods 
\cite{cooper,sinova,rozhkov} is found to be $c_{L}=(c_{L})_{m}\sim 0.1 $\,. 
The vortices are in the lattice phase if $c_{L}<(c_{L})_{m}$, otherwise in 
molten phase if $c_{L}>(c_{L})_{m}$. One could also say that the vortex 
crystal starts melting as the vortices begin to depart from thier lattice 
positions. 

However in this work, I am taking a completely diverse approach from the one
discussed above. The difference is two-fold, firstly I am already in a 2D 
geometry so that an ultra-fast rotation to achieve this geometry is not 
required. Secondly I assume my vortices are in the liquid phase. Freezing and 
strongly correlations need to be analyzed hence after. The following sections 
will shed more light on my approach to the $N$ vortex problem in the confined 
2D geometry.

\subsection{Vortices as charged bosons}

Let us consider a system of $N$ bosons in a condensate confined within a
cylindrical potential (with radial $\omega_{\perp}$ and axial trap 
frequencies $\Omega_{z}$) and put into rotation with angular velocity 
$\Omega$. The Hamiltonian of the system reads \cite{legget,sinova} 
\begin{eqnarray}
H_{\Omega}&=& \sum_{i=1}^{N} \frac{{\bf p}_{i}^2}{2m} 
+\frac{1}{2}m\omega^2 {\bf r}_{i}^2-\Omega\, \hat{z}\cdot {\bf r}_{i}
\times{\bf p}_{i}+\sum_{i<j=1}^{N}V({\bf r}_{i}-{\bf r}_{j})\nonumber \\
&=&\sum_{i=1}^{N}\frac{({\bf p}_{i}-m\Omega\hat{z}\times{\bf r}_{i})^2}{2m}
+\frac{1}{2}m[(\omega_{\perp}^2-\Omega^2)r_{\perp}^2
+\Omega_{z}^2z_{i}^2]\nonumber \\
&+& \sum_{i<j=1}^{N}V({\bf r}_{i}-{\bf r}_{j})\nonumber \,.
\label{tre-tiger}
\end{eqnarray}
This system is identical to a system of charge-$Q$ particles interacting under 
the influence of a magnetic field 
\begin{equation}
 B^{*}= \nabla \times(m\Omega\hat{z}\times{\bf r}/Q)=(2m\Omega/Q)
\hat{z}.\nonumber
\end{equation}
At sufficiently low temperatures and in the fast rotation limit 
$\Omega\gg\,ms^2$, this system enters an effectively two-dimensional strong 
field and the dissipation due to scattering of thermal excitations on the 
vortices is negligible, thus the vortex dynamics is conservative. The 
situation considered then corresponds to the Magnus-force dominated limit of 
vortex motion in a very clean superconductor \cite{blatter,kopnin}. 

An indirect connection to the above argument can be related to the work of
Nelson et al. \cite{nelson1} who have shown that the statistical mechanics of 
the flux-line lattice (FLL) of high-$T_{c}$ superconductors can be studied 
through an appropriate mapping onto the 2D Yukawa boson system. Following this 
magnificent idea, Magro and Ceperley \cite{Magro93a,Magro94a} have performed 
Diffusion Monte Carlo (DMC) and Variational Monte Carlo (VMC) numerical 
calculations to calibrate the vortex properties for a homogeneous 2D Yukawa 
boson system. Taking a similar step, I will treat the system of 
$N_{v}$ vortices  as $N$ bosons confined in a 2D harmonically planar trap
$\frac{1}{2}m\omega_{\perp}^2 r^2$  performing a Density functional theory 
(DFT) calculation (See Chaprter 3.) 

\begin{figure}
\centering{
\epsfig{file=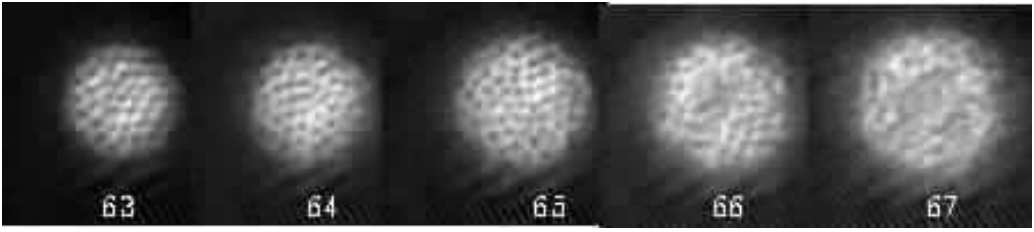,width=1.0\linewidth}}
\caption{Vortices when rotation $\Omega\rightarrow \omega_{\perp}$. 
Centrifugal anf trapping energy compensate and spatial extent becomes large. 
Adapted from Bretin et. al. \cite{bretin} }
\label{fig_lattice}
\end{figure}

\subsection{Quantum Monte Carlo Techniques (VMC and DMC)}

My DFT calculation heavily relies on the accurate data of VMC calculation
of Magro and Ceperley \cite{Magro93a}. So I feel it is neccessary to display 
here some idea of what the method is all about. VMC method is a technique 
developed to minimize the ground-state energy of a many-body trial function. 
The energy functional $E=\bra{\phi_{T}}H\ket{\phi_{T}}$, where $\phi_{T}$ is a 
family of trial functions is minimized by one of its particular choice of 
wave-function $\phi_{0}$ known as the ground state. In other word we introduce 
a family of trial function $\phi(x,p)$, depending on the parameter $p$ and 
compute the corresponding energy,
\begin{equation}
E(p)= \frac{\int\,\phi^{*}(p,r)\,H\,\phi(p,r)}{\int\,\phi^{*}(p,r)\phi(p,r)}
\label{tre-kija}
\end{equation}
and look for the choice of $p=p*$ such that $E^{*}$ is the minimum energy 
within the parameter space. Obviously the success of VMC depends largely on 
a good choice of trial function. The most commonly used trial function when 
one is interested in identifying the solid or liquid phases of a system is the 
Jastrow pair product which reads \cite{Magro93a}
\begin{equation}
\phi_{T}=\prod_{i<j}\exp[-u(r_{ij})]\prod_{i}
\exp[-c({\bf r}_{i}-{\bf Z}_{i})^2]
\label{tre-ular}
\end{equation}
where ${\bf r}_i$ is the position of $i\,th$ of $N$ particles, ${\bf Z}_{i}$ 
are the lattice sites, $r_{ij}$ is an interparticle distance, and $u(r_{ij})$ 
is the Jastrow pair function chosen to approximate the solution of two-body 
Schr{\"o}dinger equation. The lattice is assumed to be triangular since this 
minimizes the potential energy \cite{abrikosov2}. The Jastrow function 
contains parameters which along with $c$ are varied to 
minimize $E(p)$. For the liquid phases $c=0$ but for a solid $c$ takes a 
finite value treated as an adjustable parameter. Once a right choice of 
variational function is obtained, the ground state energy is further 
calculated using the DMC method. Starting with the imaginary time 
Sch{\"o}dinger equation,
\begin{equation}
-\frac{\partial \psi(R,t)}{\partial t}=(H-E_{T})\psi(R,t)
\label{tre-tikus}
\end{equation}   
and by introducing $f(R,T)=\phi_{T}(R)\psi(R,t)$ as the mixed distribution 
function to the equation above we can obtain the diffusion-reaction 
equation \cite{Magro93a},
\begin{equation}
\frac{\partial f}{\partial t}=\Lambda\sum_{i}\nabla_{i}^2 f
-2\Lambda^2\sum_{i}\nabla_{i}\cdot(f\nabla_{i}\phi_{T})-(E_{L}-E_{T})f
\label{tre-anac}
\end{equation}
where $\Lambda$ is the dimensionless DeBoer parameter defined as 
$\Lambda=\hbar^2/2m\sigma^2\epsilon$  in which $\sigma$ and $\epsilon$ are the 
reduced unit length and energy. $E_{L}=H\phi_{T}/\phi_{T}$ found in the last 
term represents the upper bound on the ground state energy. The value of 
$E_{T}$ that gives a steady poulation is the exact ground state energy. Other 
properties such as the pair distribution functions are also calculated from 
the mixed distribution, $\phi_{T}(R)\psi_{0}(R)$ where $\psi_{0}(R)$ being the 
ground state in a large time evolution. The pair-correlation function $g(r)$ 
and the structure function $S(k)$ encode the information on the degree of 
order of the system. Oscillating pair-correlation function is what one 
anticipates when the system is in solid phase while a non-oscillating one 
signatures a liquid phase.

\section{Brief outline of the thesis}

The microscopic formalism developed in the first part of this thesis can 
accurately treat a weakly interacting Bose-Einstein condensate and 
a non-condensed thermal atoms for a range of temperatures below the 2D 
critical transition temperature $T_{c}$. In the same spirit, I assume the 
semi-classical Two-fluid model to be valid in evaluating the equilibrium 
property of the condensate fraction and the density profile of bosonic gas 
in this limit. In this model the density profiles are calculated 
self-consistently by ensuring that the choice of effective interaction 
potential is introduced in a consistent manner. It is an extremely important 
point since the introduction of inaccurate effective potential costs 
heavily in the convergence of the numerical solution. Another crucial fact 
to consider is the scattering property of a strictly 2D Bose gas, where 
the $s$-wave scattering length has become larger than the axial thickness of 
the cloud changes dramatically. In Sec. 2.1 I will illustrate how  a 
semi-analytical expression for the many-body T-matrix elements corresponding 
to the condensate-condensate and to the condensate-thermal cloud coupling 
parameters ($g_{2}$ and $g_{1}$, respectively) effects the equilibrium 
properties of the Two-fluid model at increasing temperature.

Sec. 2.2  summarizes the Feshbach formalism framework with application to the 
2D Bose gas. Here I examine the tuning of two-body collisions by 
means of a magnetic field in a strictly 2D condensate inside a harmonic trap 
at zero temperature. By modifying the coupling strength one can control 
the dimensionless physical parameter $\eta$ which is the ratio between the 
mean-field interaction energy and the harmonic-oscillator level 
spacing \cite{kagan2}. In the case of repulsive interactions ($\eta>0$),
starting from a weakly interacting regime ($\eta\ll1$) and up to a strongly 
interacting one ($\eta\gg 1$), I evaluate the equilibrium density profile as 
a function of $\eta$ by solving numerically a non-linear Schr\"odinger 
equation by a split-step Crank-Nicholson discretization scheme. In the case of 
attractive interactions ($\eta<0$) one can give an upper bound for the number 
of bosons which can condense without the condensate collapsing. At variance 
from the 3D gas (see for instance \cite{antonius,savage}), I have found that 
in 2D this critical number does not depend explicitly on the strength of the 
harmonic confinement.

In Sec. 3.1, I consider a 2D rotating condensate at finite temperature and 
evaluate the particle density profiles and the energy of a vortex within a 
strictly 2D model for the boson-boson coupling. This is appropriate to a 
situation in which the {\it s}-wave scattering length starts to exceed the 
vertical confinement length. Previous work has established that the 
dimensionality of the scattering collisions strongly affects the equilibrium 
density profiles \cite{tanatar,rajagopal} and the process of free expansion of 
a BEC containing a vortex \cite{hosten}. A similar study of the density 
profiles of a rotating BEC in 3D geometry at finite temperature has been 
carried out by Mizushima {\it et al.} \cite{mizushima}, who also determined 
the location of various dynamical instabilities within the Bogoliubov-Popov 
theory.

In Sec. 3.2, I will report the result of my investigation on the ground state 
properties of $N_{v}$ vortices treated as $N$ Yukawa bosons trapped in the 
two-dimensional harmonic trap. A density functional theory (DFT) approach is 
developed to calculate the density profiles and the ground state energy of the 
vortex fluid within the Bogoliubov-de Gennes reference system. When a 
non-interacting auxiliary system is inferred in this model, we obtained a 
Kohn-Sham (KS) type equation. As a preliminary step, the density profile of 
the vortex fluid is calculated by solving this equation numerically. The 
method relies on the VMC data of Magro and Ceperley \cite{Magro93a} on the 
homogeneous ground state energies. The work in this thesis also pave way to 
many unresolved open questions pertaining to the physics of freezing and that 
of the strongly correlated incompressible liquid which has an overlap with the 
Fractional Quantum Hall Effect (FQHE).

%\section*{Introduzione}
\clearpage{\pagestyle{empty}\cleardoublepage} 
% se c'e' la pagina bianca fa lo stile empty

%%%%%%%%%%%%%%%%%%%%%%%%%%%%%%%%%%%%%%%%%%%%%%%%%%%%%%%%%%%

%\clearpage{\pagestyle{empty}\cleardoublepage}
\chapter[Scattering properties for 2D Bose gas]{Scattering properties for 2D Bose gas}
\label{shuttle_chapter}
  \section{Coupling parameters in a mixture of condensed and thermal-bosons}
  \label{sec_T}

   The scattering property of bosons in a 2D flat geometry changes 
   dramatically compared to the 3D one. The T-matrix for two-body 
   collisions {\it in vacuo} at low momenta and energy, which should be used 
   to obtain the collisional coupling parameter to the lowest order in the 
   particle density, vanishes in the strictly 2D limit \cite{popov,schick}. 
   It is then  necessary to evaluate the scattering processes between pairs 
   of Bose particles by taking into account the presence of a condensate and 
   a thermal cloud through a many-body T-matrix formalism 
   \cite{stoof93,bijlsma97,Lee}. 
   
   This formalism has already been used in a number of studies of low 
   dimensional Bose gases, dealing in particular with phase fluctuations and 
   the Kosterlitz-Thouless transition in a variational approach 
   \cite{stoof93,bijlsma97,khawaja,khawaja2}, with a mean-field evaluation of 
   the breathing-mode frequency in a trapped 1D gas \cite{proukakis}, and with 
   the equilibrium density profile of a 2D condensate \cite{Lee,tanatar}. In 
   the limit of zero temperature the condensate-condensate coupling parameter 
   has been related to the two-body T-matrix by considering that, when two 
   particles in the condensate collide at zero momentum, they both require an 
   energy equal to the chemical potential  $\mu$ to be excited out of the 
   condensate \cite{Lee}. With increasing temperature the population of the 
   excited states becomes non-negligible and a microscopic theory which also 
   takes into account the depletion of the condensate is required 
   \cite{proukakis2}. An additional coupling parameter to describe the 
   scattering processes between an atom in the condensate and an atom in the 
   thermal cloud has been introduced by Stoof and co-workers \cite{khawaja} 
   through the two-body T-matrix at energy $-\mu$.
 
  The scattering processes between pairs of atoms in a gas consisting of a 
  Bose-Einstein condensate cloud and a thermal cloud are described by the 
  many-body T-matrix $T^{MB}(E)$ as a function of the energy $E$. The coupling
  parameters are given by the matrix elements 
  $\bra{{\bf k'}}T^{MB}(E)\ket{{\bf k}}\equiv T^{MB}({\bf k},{\bf k'},{\bf K};E)$, taken in the limit of zero energy and momenta. 
  Here ${\bf k}$ and ${\bf k'}$ are the incoming and outgoing relative momenta 
  of the pair of center-of-mass momentum ${\bf K}$. I will consider only the 
  condensate-condensate and condensate-thermal cloud couplings and neglect the 
  scattering between thermally excited atoms in the present case of a dilute 
   gas. 

  Before discussing the many-body T-matrix, however, we shall first recall the
  behaviour of the two-body T-matrix that describes collisions between pairs 
  of particles {\it in vacuo}. We shall then discuss how the two T-matrices 
  are related in the appropriate limit for a two-fluid system. 

  \subsection{The two-body T-matrix}

  The two-body T-matrix is the solution of the Lippmann-Schwinger equation,
  \begin{eqnarray}
  \bra{\bf k'}T^{2B}(\bar{E})\ket{{\bf k}}&=&\bra{{\bf k'}}V(|{\bf r}_{1}-
   {\bf r}_{2}|)\ket{{\bf k}}
   +\sum_{{\bf q}}\bra{{\bf k'}}V(|{\bf r}_{1}-{\bf r}_{2}|)\ket{{\bf q}} 
   \nonumber \\ 
   &&\times \frac{1}{\bar{E}-2\epsilon_{\bf q}}\bra{{\bf q}}T^{2B}(\bar{E})
   \ket{{\bf k}}\,,
  \label{two-amma}
   \end{eqnarray}
   with $V(|{\bf r}_{1}-{\bf r}_{2}|)$  being the interparticle potential. The 
   center-of-mass energy for the pair of atoms is $\bar{E}$ and each atom of mass
   $m$ has a single-particle excitation energy $\epsilon_{{\bf q}}=\hbar^2 q^2/2m$, 
   as the collision takes place in free space. In Eq. (\ref{two-amma}) the 
   collisions are described by a single-loop interaction between the two atoms plus 
   contributions involving all possible transition routes from state  
   $\ket{{\bf k}}$ to state $\ket{{\bf k'}}$ {\it{via}} intermediate states 
   $\ket{{\bf q}}$.
 
   In the case of the hard-disk potential of strength $V_{0}=4\pi\hbar^2/m$ and in
   the dilute limit (${\bf k}a, {\bf k'}a\ll 1$), the solution of 
   Eq. (\ref{two-amma}) is \cite{khawaja}
   \begin{equation}
    \bra{{\bf k'}}T^{2B}(\bar{E})\ket{{\bf k}}
   \approx \frac{4\pi\hbar^2/m}{\ln|4\hbar^2/\bar{E}ma^{2}|}. 
   \label{two-appa}
   \end{equation}
   In this case the T-matrix is independent of the momenta, but depends on the
   logarithm of the energy and vanishes as $\bar{E}$ approaches zero.
   Therefore, the presence of the surrounding gas must be taken into account in 
   the collisional processes. Following the proposal of Morgan et al. 
   \cite{morgan}, this is done by setting $\bar{E}=-2\mu$ for the mutual scattering 
   of two condensate particles, this being the energy required for them to be 
   excited out of the condensate. By a similar argument Al Khawaja et al. 
   \cite{khawaja} set $\bar{E}=-\mu$ for the scattering between a boson in the 
   condensate and boson in the thermal cloud.

   \subsection{The many-body T-matrix}
    
   According to the above argument, a first approximation for the many-body 
   T-matrix at zero momenta and energy is
   \begin{equation}
   T_{n}^{MB}(0,0,0;0)=T^{2B}(0,0,0;-n\mu)
   \label{two-tata}
   \end{equation}
   with $n$ = 1 for condensate-thermal cloud scattering and $n$ = 2 for
   condensate-condensate scattering. Further many-body effects can enter the
   scattering processes attended by the presence of a condensate from attributing a
   Bogoliubov spectrum to the intermediate states \cite{stoof93}. At low momenta
   the many-body T-matrix is the solution of the integral equation
   \begin{eqnarray}
   \bra{\bf{k'}}T^{MB}(E)\ket{\bf{k}}&=&\bra{\bf{k'}}V(|\bf{r}_{1}-\bf{r}_{2}|)
   \ket{\bf{k}}+\sum_{\bf{q}}\bra{\bf{k'}}V(|\bf{r}_{1}-\bf{r}_{2}|)\ket{\bf{q}} 
   \nonumber\\ 
   && \times \frac{1+2N(\omega_{\bf{q}})}{E-2\hbar \omega_{\bf{q}}}
   \bra{\bf{q}}T^{MB}(E)\ket{\bf{k}}
   \label{two-paddi}
   \end{eqnarray}
   where $\hbar \omega_{{\bf q}}\approx \epsilon_{{\bf q}}+\mu$ are the Bogoliubov 
   excitation energies  in the Hartree approximation $\mu\ll\epsilon_{{\bf q}}$ and 
   $ N(\omega_{{\bf q}})\approx {\exp[\beta(\epsilon_{{\bf q}} +\mu)]-1}^{-1}$ are 
   the corresponding population factors. For a contact interaction potential of 
   Eq. (\ref{two-paddi}) yields
   \begin{equation}
    T^{MB}(0,0,0;E) =\left [ \frac{1}{V_{0}}- \sum_{\bf{q}}
    \frac{1+2N(\omega_{\bf{q}})}{E-2\hbar \omega_{\bf{q}}}\right]^{-1}.
   \label{two-mami}
   \end{equation}
   As discussed by Stoof and Bijlsma \cite{stoof93,bijlsma97}. Eq. (\ref{two-mami}) 
   is still affected by an ultraviolet divergence, which can be remedied by 
   replacing $V_{0}$ in favor of the two-body T-matrix. The final result is
   \begin{equation}
    T^{MB}_{n}(0,0,0;0) =T^{2B}(0,0,0;-n\mu)\left [ 1 + T^{2B}(0,0,0;-n\mu)
    \sum_{\bf{q}}\frac{N(\omega_{\bf{q}})}{\hbar \omega_{\bf{q}}}\right]^{-1},
   \label{two-macci}
   \end{equation}   
   with $T^{2B}(0,0,0;-n\mu)$ given by Eq.(\ref{two-appa}) with $\bar{E}=-n\mu$.
 
   \subsection{Calculation of coupling parameters}

\begin{figure}
   \centering{
   \epsfig{file=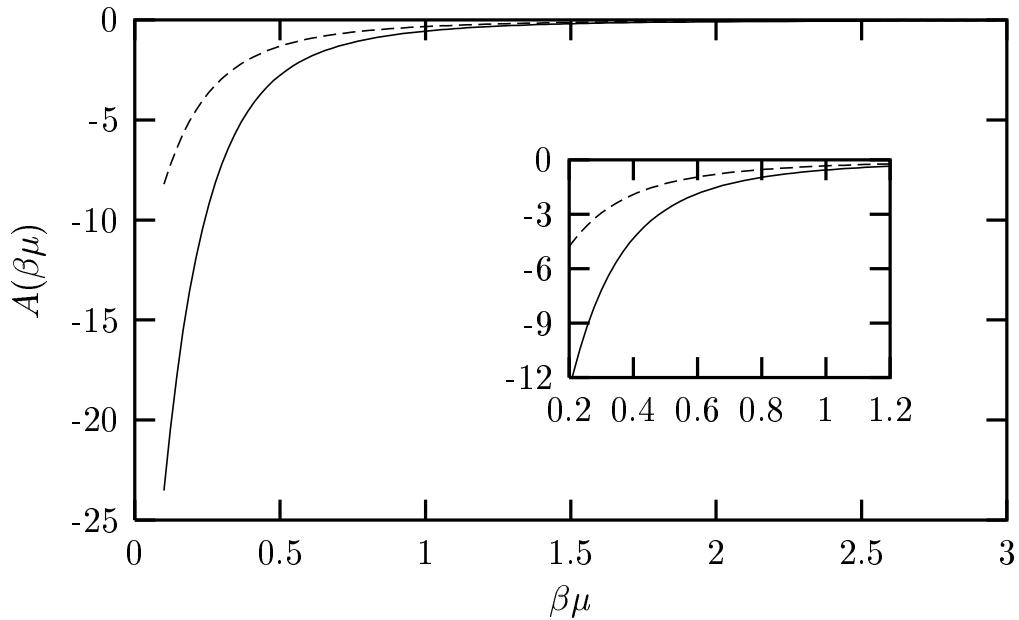,width=1.0\linewidth}}
   \caption{The correction term $A(\beta\mu)$=$\sum_{s=1}^{\infty}Ei(-s\beta\mu)$
    from excited-state occupancy (dashed line) and its approximate form 
    Eq. (\ref{two-reza})
    (solid line) as functions of $\beta\mu$. In the inset a zoom of the region
     $0.2<\beta\mu<1.2$ is shown.}
    \label{two-figcorr}
    \end{figure}

   The sum in the RHS of Eq.(\ref{two-macci}) can be evaluated analytically by 
   replacing the sum over intermediate states by an integral over momentum. 
   Setting $N(\omega_{{\bf q}})=
   \sum_{s=1}^{\infty}\exp(-s\beta\hbar\omega_{\bf{q}})$, we get 
   \begin{equation}
   \sum_{q}\frac{N(\omega_{q})}{\hbar\omega_{q}}= -\frac{m}{2\pi \hbar^2}\sum_{s=1}^{\infty} 
    Ei (-s\beta\mu)
   \label{two-pavel}
   \end{equation}
   where $Ei$($x$) is the exponent-integral function. In the asymptotic 
   low-temperature regime ($\beta\mu\gg1$) this thermal-population term can be 
   approximated by
   \begin{equation}
   \sum_{s=1}^{\infty} Ei (-s\beta\mu)\rightarrow
     (\beta \mu)^{-1}\ln[1-\exp(-\beta\mu)].
   \label{two-reza}
   \end{equation} 
   A numerical illustration is given in Fig. ({\ref{two-figcorr}}) for values 
   of the system parameters appropriate to the experiments on $^{23}$Na by 
   G\"orlitz {\it et al.} \cite{gorlitz} (see Sec. (\ref{sec-TFequil})).

    In summary, the coupling parameters in the 2D Bose gas are given by
    \begin{equation}
    g_{n}^{MB}= \frac{4\pi\hbar^2/m}{\ln|4\hbar^2/(nm\mu a^{2})|-2
    \sum_{s=1}^{\infty}Ei (-s\beta\mu)},
   \label{two-mbcoup}
   \end{equation}
   with $n$=2 for collision between pairs of condensate atoms and $n$=1 for 
   collisions between an atom in the condensate and thermally excited atom. 
   These parameters depend on temperature both through the chemical potential 
   and through the excited-state population factor given in 
   Eq. (\ref{two-pavel}) and asymptotically approximated by 
   Eq.(\ref{two-reza}). If this population factor is dropped, one obtains from 
   Eq.(\ref{two-tata}) the corresponding ``two-body'' coupling parameters as 
  \begin{equation}
    g_{n}^{2B}= \frac{4\pi\hbar^2/m}{\ln|4\hbar^2/(nm\mu a^{2})|}\,.
   \label{two-2bcoup}
   \end{equation}
   In the next Section I evaluate the chemical potential and hence the 
   coupling parameters in Eqs. (\ref{two-mbcoup}) and (\ref{two-2bcoup}) 
   through a self-consistent 
   evaluation of the density profiles in a two fluid-model.

   \subsection{Equilibrium properties in the two-fluid model}
   \label{sec-TFequil}

   Let $n_{c}(r)$ and $n_{nc}(r)$ be the particle density profiles for the
   condensate and for the thermal cloud in a 2D Bose gas which is radially 
   confined inside a isotropic planar trap described by the external potential
   $ V_{ext}(r)=\frac{m \omega_{\perp}^2 r^2}{2}$. The two-fluid model 
   \cite{minguzzi97,vignolo} combines a solution of the Gross-Pitaevskii 
   equation for the condensate with a Hartree-Fock model of the thermal cloud, 
   which is treated as an ideal gas subject to an effective potential 
   $V_{eff}(r)$. In the present case, 
   \begin{equation}
   V_{eff}(r)= V_{ext}(r) + 2g_{1}n_{c} .
   \label{two-veff}
   \end{equation} 
    As already noted, we are neglecting the collisions between pairs of bosons
    belonging to the thermal cloud.

    It is well known that a full numerical solution of the Gross-Pitaevskii 
    equation Eq. ({\ref{one-GP}}) can be avoided when the kinetic energy 
    term in it can be neglected \cite{baym96}. I postpone the evaluation of a 
    full scale numerical solution of the GPE equation for the coming 
    Sec. (\ref{Sec_fesh}) on Feshbach resonances and in chapter 3, where the 
    role of the kinetic energy become crucial in vortex calculation. Since our 
    goal here is to analyse the effect of the interactions on the equilibrium 
    properties of the system, we use the Thomas-Fermi approximation in 
    which kinetic energy is neglected. This yields a condensate density of 
    the form
   \begin{equation}
   n_{c}(r)= (1/g_{2})[ \mu - V_{ext}(r) - 2g_{1}n_{nc}(r) ]\theta(\mu - 
   V_{ext}(r) - 2g_{1}n_{nc}(r)).
   \label{two-Cden}
   \end{equation}
   where $\theta(x)$= 0 for $x<0$ otherwise $\theta(x)$= 1 if $x>0$. 
   Since I am interested in examining qualitative behaviours rather than in 
   attaining high numerical accuracy, I have adopted Eq.(\ref{two-Cden}) 
   for the condensate density and found that discontinuities occurring in the 
   density profiles at the Thomas-Fermi radius can be eliminated by the simple 
   expedient of introducing a momentum cut-off in the expression for 
   $n_{nc}(r)$ of Eq. (\ref{one-venty6}). That is, we calculate $n_{nc}(r)$ 
   from   
  \begin{equation}
   n_{nc}(r)=-\frac{m}{2 \pi \hbar^2\beta} \ln \left[1- exp[\beta(\mu - 
   V_{eff}(r)-
   \frac{p_{0}^2}{m})]\right],
   \label{two-Tden}
   \end{equation}
   where we take $p_0=\sqrt{mg_{1}n_{nc}}$  \cite{prokof}. The model is then 
   evaluated by solving self-consistently 
   Eqs. (\ref{two-veff})-(\ref{two-Tden}) together with the condition that the 
   areal integral of $n_{c}(r)$+$n_{nc}(r)$ should be equal to the total 
   number $N$ of particles.

  The validity of the present model is as follows. A mean-field treatment is 
  valid when the diluteness condition  $n_{c}a^2\ll 1$ holds and if the 
  temperature of the gas is outside the critical region. With regard to the 
  thermal cloud, no significant differences have been found between the 
  predictions of the Hartree-Fock and Popov approximations in the
  regime $n_{c}a^2\ll 1$, except at very low temperature where the thermal 
  cloud is becoming negligible \cite{dalfovo}.
    
   \subsection{Coupling strength and its effect on equilibrium properties}
   \label{two-conclu}
 
   For a numerical illustration, we have taken the values of particle number, 
   the radial trap frequency, and the scattering length as appropriate for
   $^{23}$Na atoms in the experiment of G\"orlitz {\it et al.}\cite{gorlitz} 
   ($N =5\times 10^5,\,\omega_{\perp}=188.4\,$Hz, $a=2.8$ nm). Whereas in 
   their experiment collisions are in 3D regime, we focus on a strictly 2D 
   regime that could be reached experimentally by increasing either the trap 
   anisotropy parameter or the scattering length.

\begin{figure}
   \centering{
   \epsfig{file=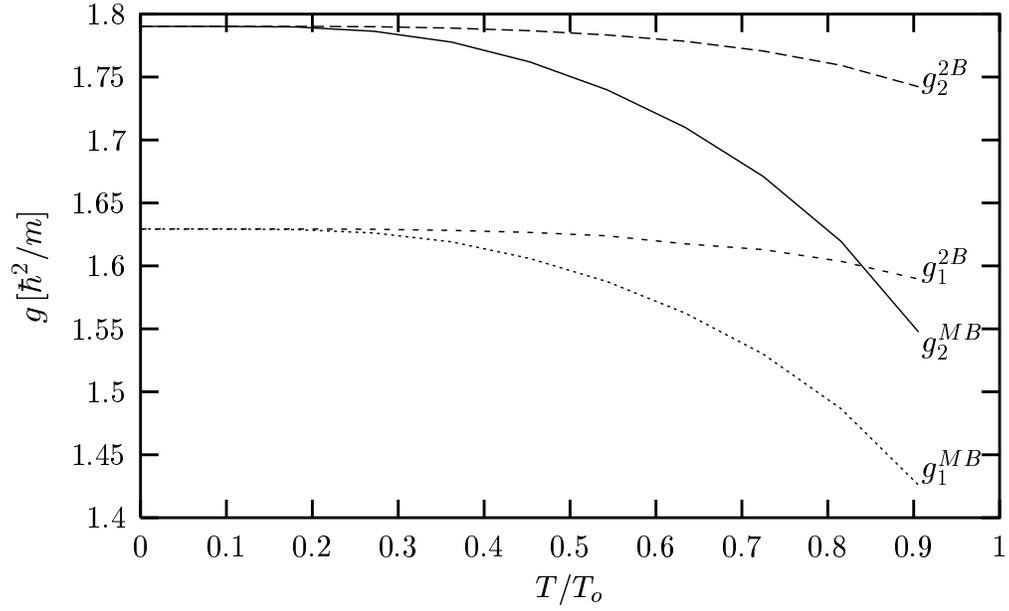,width=1.0\linewidth}}
   \caption{Interaction strengths (in units of $\hbar^2/m$) as functions of 
    temperature $T$ (in units of ideal-gas critical temperature $T_{0}$).}
   \label{two-figcoup}
   \end{figure}

   I have compared the temperature dependence of the coupling parameters 
   $g_{n}^{2B}$ (long dashes for $n$=2 and short dashes for $n$=1) with that 
   of the coupling parameters $g_{n}^{MB}$ (full and dotted lines, 
   respectively) in Fig. (\ref{two-figcoup}). It is evident that the many-body 
   screening of the interactions due to the occupancy of excited states is 
   quite large and rapidly increasing with temperature.

\begin{figure}
    \centering{
    \epsfig{file=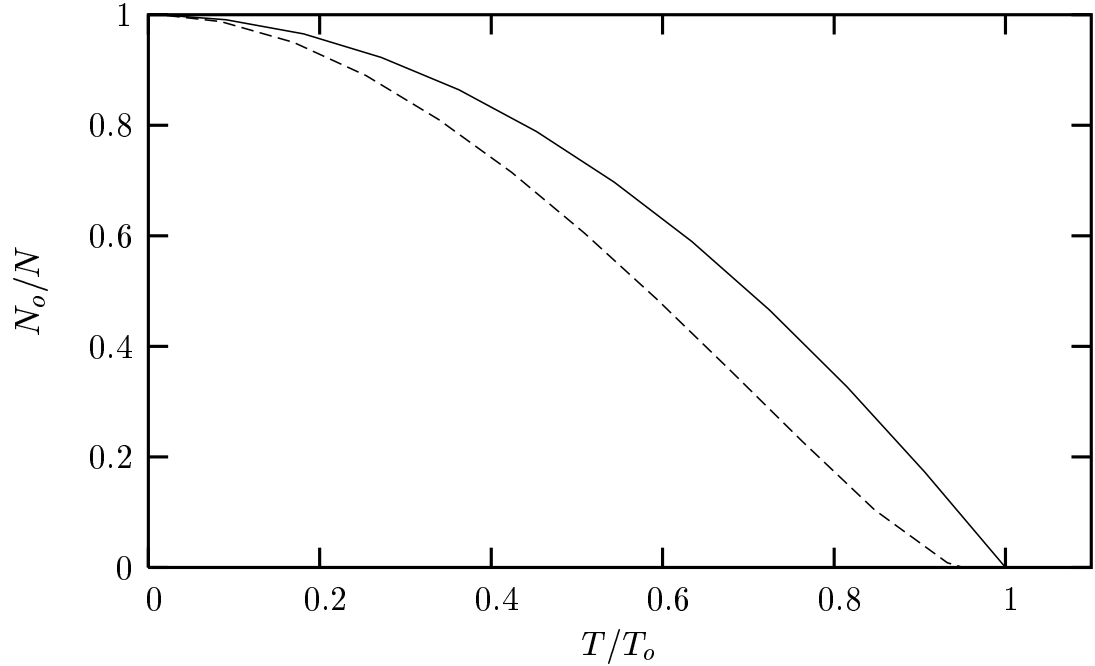,width=1.0\linewidth}}
    \caption{Condensate fraction $N_{0}/N$ as a function of temperature $T$
    (in units of $T_{0}$) from the $g_{n}^{MB}$ model and (dashed line) 
     compared with the non-interacting gas (solid line).}
    \label{two-figfrac}
     \end{figure}

   Such many-body screening has, however, very little effect on equillibrium 
   properties of the gas for our choice of system parameters. 
   Figure (\ref{two-figfrac}) reports the condensate fraction $N_{0}/N$ for 
   the $g_{n}^{M}$ model as a function of temperature, in comparison with that 
   of an ideal Bose gas at the same values of the system parameters. As is 
   well known, the transition temperature and the condensate fraction are 
   lowered by the interactions. However, the $g_{n}^{2B}$ model gives results 
   that are practically the same as the $g_{n}^{MB}$ one.\par   

\begin{figure}
   \centering{
   \subfigure[]{
   \epsfig{file=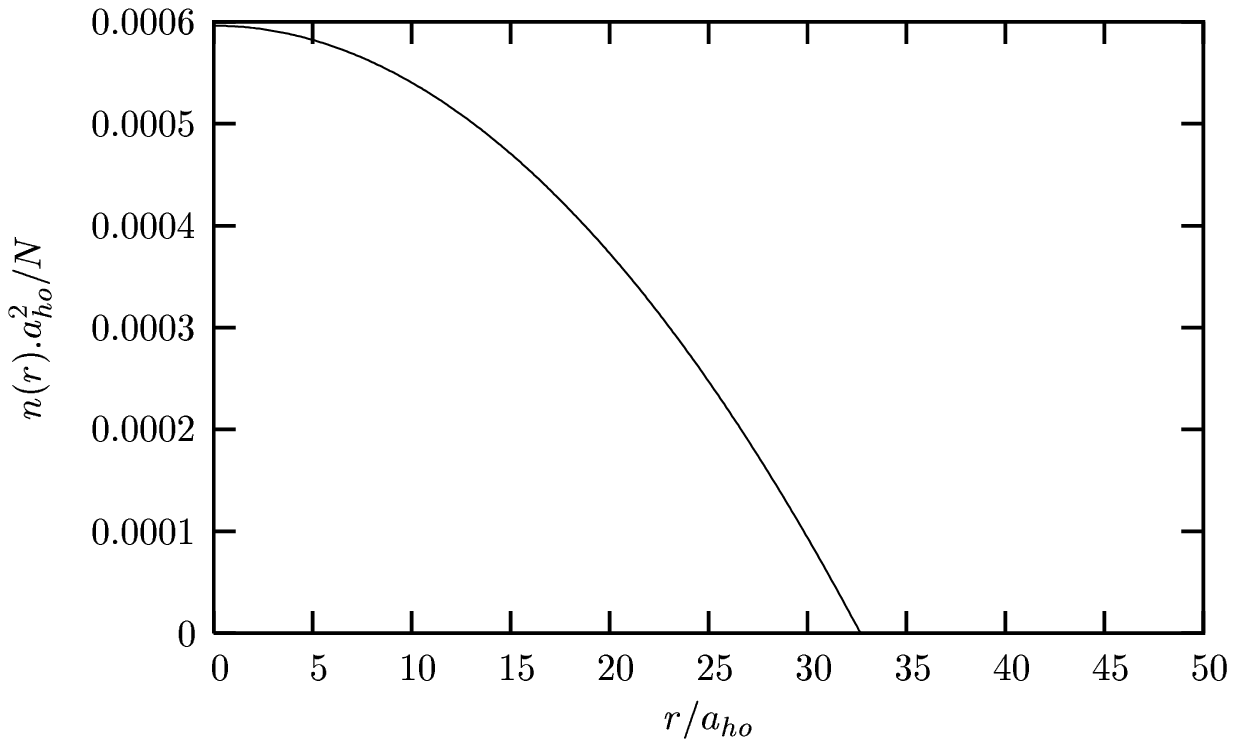,width=0.46\linewidth}}
   \subfigure[]{
   \epsfig{file=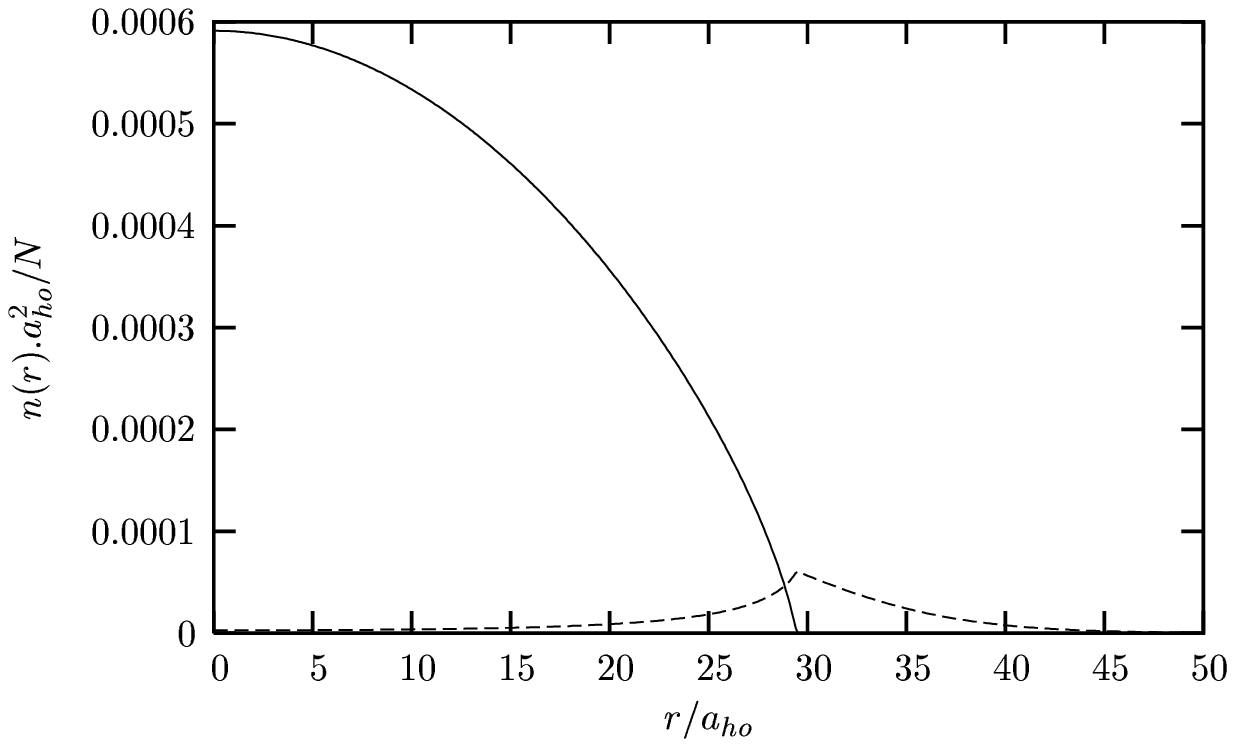,width=0.46\linewidth}}
   \subfigure[]{
   \epsfig{file=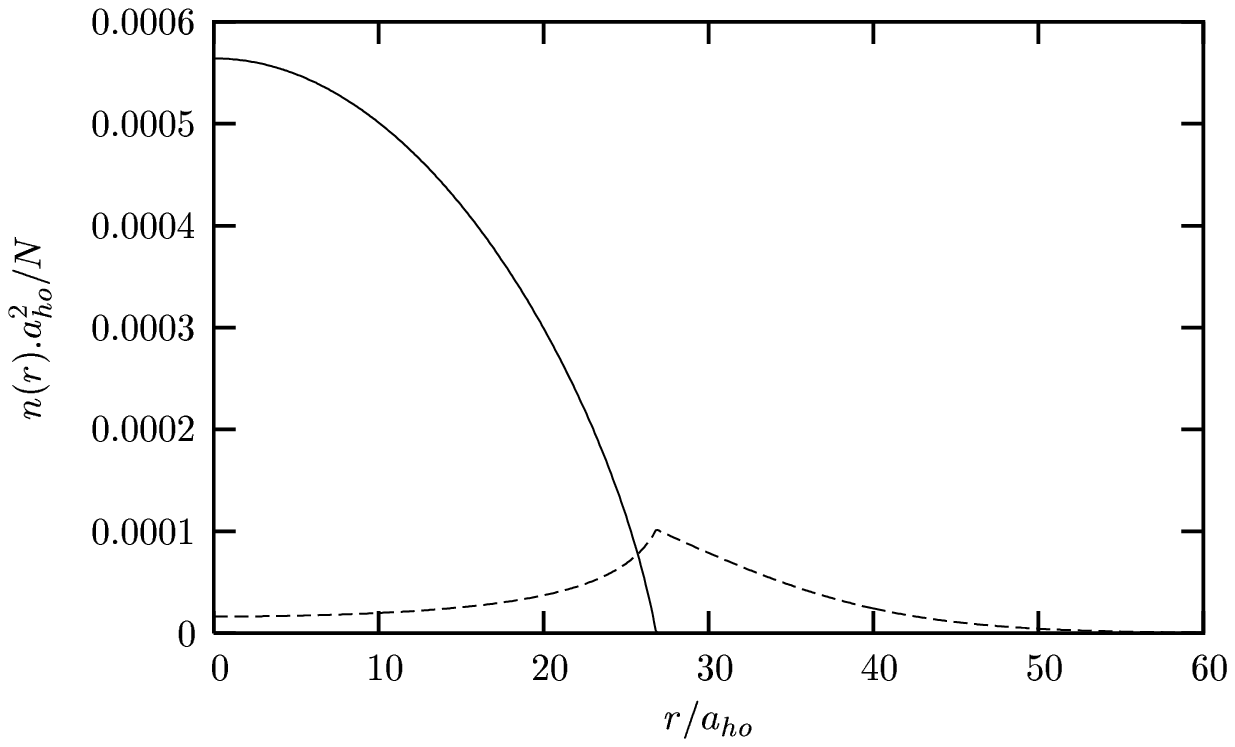,width=0.46\linewidth}}
   \subfigure[]{
   \epsfig{file=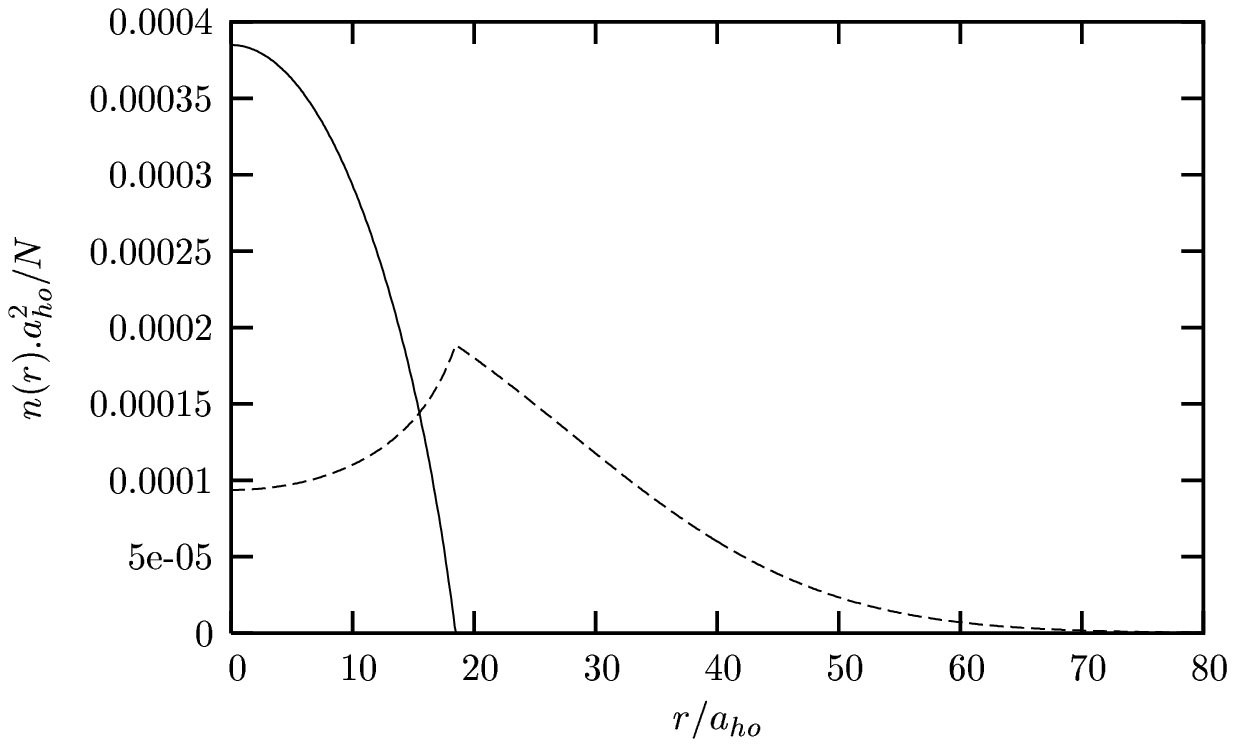,width=0.46\linewidth}}
   \subfigure[]{
   \epsfig{file=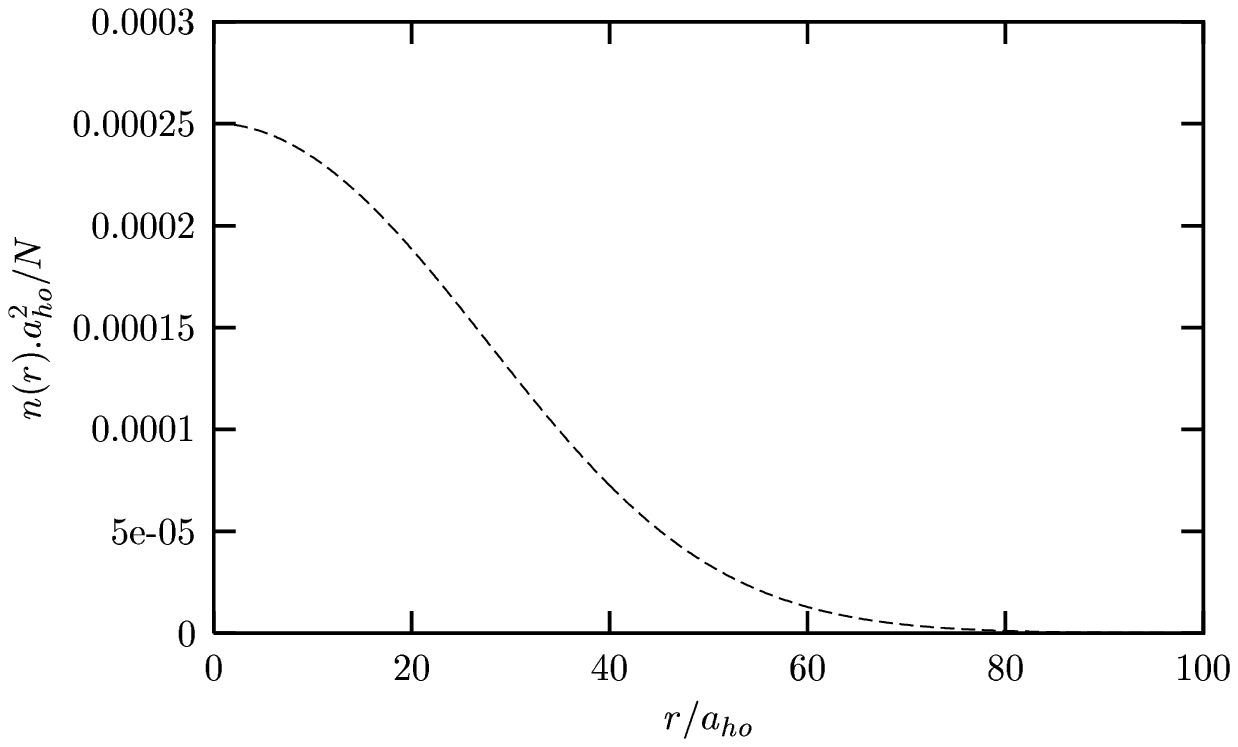,width=0.46\linewidth}}
   }
   \caption{Density profiles of the condensate (solid line) and 
   the thermal cloud (dashed line) in  the $g_{n}^{MB}$ model (in units of 
   $N/a^2_{ho}$ with $a_{ho}=(\hbar/m\omega)^{1/2}$) as functions of radial 
   distance $r$ (in units of $a_{ho}$ ) at various values of the 
   temperature  ( $T/T_{0}$=0,0.3,0.5,0.8 and 0.95 from (a) to (e)).}
   \label{two-figevol}
   \end{figure}

   The evolution of the density profiles for the condensate and for the 
   thermal cloud with increasing temperature from near absolute zero to the 
   critical temperature $T_{0}$ of the ideal gas 
   ($T_{0}=(\sqrt{6}/\pi)\hbar\omega\sqrt{N}$ \cite{bagnato}) are depicted
   in Figure (\ref{two-figevol}). The results in Figure (\ref{two-figevol}a) 
   are in good agreement with those of Tanatar et al. \cite{tanatar}, except 
   that the tails of that profile are missed in the Thomas-Fermi 
   approximation. Again the screening of collision from the occupancy of the 
   excited states is very small and becomes barely visible at 
   $T \approx$ 0.75 $T_{0}$.

   We can deduce that the use of Eq. (\ref{two-tata}) to describe the 
   many-body effects in two-body scattering processes in 2D Bose-condensed gas 
   appears to be very good in regard to equilibrium properties at large 
   values of the particle number. A decrease in the number of particles lowers 
   the chemical potential and may lead to observable effects for 
   $N\approx$ 10$^{3}$.  

   \newpage

   \section{Feshbach resonance in a 2D Bose gas}
   \label{Sec_fesh}

     The 3D $s$-wave scattering length can be tuned by means of external 
     electric or magnetic fields \cite{tiesinga93,moerdijk95,vogels}. Within 
     this process a resonance can appear at a particular value of the magnetic 
     field and allows control of the coupling strength ranging from positive 
     (repulsive) to negative (attractive) values. One may study new dynamical 
     effects of collective oscillations in large condensates \cite{kagan} 
     and the dynamics of collapsing and exploding condensates \cite{donley}.
     Calculations of such Feshbach resonances have been given for 3D gases 
     of $^7$Li \cite{fedichev}, $^{23}$Na \cite{moerdijk95}, $^{39}$K and 
     $^{41}$K \cite{boesten}, $^{85}$Rb and $^{87}$Rb  \cite{vogels}, and 
     $^{133}$Cs \cite{tiesinga93}. Experimental observations have been 
     reported for $^{87}$Rb \cite{newbury}, $^{85}$Rb 
     \cite{donley,roberts,courteille2,cornish}, $^{23}$Na \cite{inouye}, and 
      $^{133}$Cs \cite{vuletic}. 
     Feshbach resonances in quasi-2D atomic gases have been discussed by 
     Wouters {\it et al.} \cite{wouters}, who have predicted a shift in the 
     position of the resonance from the squeezing of the confinement.

   \subsection{Overview on scattering theory}

    We start by recalling some well-known facts about a collision between two 
    particles with internal degrees of freedom governed by an internal 
    Hamiltonian $H^{int}$, 
    \begin{equation}
    H^{int}\ket{\alpha}=\epsilon_{\alpha}\ket{\alpha}
    \label{due-eins}
    \end{equation}
    where $\ket{\alpha}$ and $\epsilon_{\alpha}$ are the single-atom
    eigenstate and its corresponding eigenvalue. For example the internal 
    Hamiltonian can describe the hyperfine interaction between nuclear 
    and electronic spins and their Zeeman coupling to an external magnetic 
    field $B$. A two-body collision is described by the Hamiltonian
    \begin{equation}
    H=\frac{{ p}^2}{2 m_{r}}+\sum_{i=1}^{2}H_{i}^{int}+V({\bf r})
    \label{due-zwei}
    \end{equation}
    in the center-of-mass system, with ${\bf p}$ the relative momentum and 
    $ m_{r}$ the reduced mass (see for instance \cite{stoof88}). The 
    Hamiltonian in Eq. (\ref{due-zwei}) is the sum of a part with eigenstates
    $\ket{{\alpha\beta}}$ (``channels'') tending asymptotically to a
    symmetrized product of separate-atom internal states $\ket{\alpha}$ and 
    $\ket{\beta}$, and of a finite-range interaction $V({\bf r})$ which 
    couples the channels. The energy associated with an eigenstate 
    $\ket{{\alpha\beta}}$ is 
    $E_{\alpha,\beta}=\varepsilon_\alpha+\varepsilon_\beta+
    \hbar^2k^2_{\alpha,\beta}/2m_r$, where  $\hbar{\bf k}_{\alpha,\beta}$ is 
    the relative momentum of the two incoming particles. The asymptotic 
    scattering state $\Psi_{\alpha\beta}({\bf r})$ can be expanded on all 
    scattering channels in the form
    \begin{equation}
    \Psi_{\alpha\beta}({\bf r})=\exp({i{\bf k}_{\alpha\beta}\cdot {\bf r}})
     \ket{\alpha\beta}
     + \sum_{\alpha',\beta'} f_{\alpha\beta}^{\alpha'\beta'}
     ( {\bf k}_{\alpha\beta},{\bf k}'_{\alpha'\beta'} )
     \frac{\exp({i {\bf k}'_{\alpha'\beta'}\cdot{\bf r}})}{r}
     \ket{\alpha'\beta'}\,,
    \label{due-drei}
    \end{equation}
    where the outgoing channel is denoted by $\ket{\alpha'\beta'}$ and the 
    scattering amplitude $f_{\alpha\beta}^{\alpha'\beta'}$ is proportional to 
    the T-matrix element 
    $\bra{{\bf k}'_{\alpha'\beta'}}T \ket{{\bf k}_{\alpha\beta}}$. The 
    asymptotic magnitude of the momentum for the $\ket{\alpha'\beta'}$ channel 
    is $k_{\alpha'\beta'}=\sqrt{2 m_{r}(E_{\alpha\beta}-\epsilon_{\alpha'}-
    \epsilon_{\beta'}})$, from conservation of total momentum and energy in 
    the scattering process. The channel is said to be open (closed) if the 
    momentum takes positive real (imaginary) value.

    The whole Hilbert space describing the spatial and spin degrees of freedom 
    is now divided into a subspace P for open channels and a subspace Q 
    including all closed channels \cite{feshbach}. The state vector 
    $\ket{\Psi}$ and the Sch\"odinger equation $H\ket{\Psi}=E\ket{\Psi}$ are 
    projected onto the two subspaces by means of projection operators P and Q 
    satisfying the conditions P + Q = 1 and PQ = 0. The formal solution of the 
    coupled equations obeyed by the two projected components is
    \begin{equation}
    \ket{\Psi_{Q}}=(E-H_{QQ}+i\delta)^{-1}H_{QP}\ket{\Psi_{P}}
    \label{due-vier}
    \end{equation}
    and
    \begin{equation}
    (E-H_{PP}-H'_{PP})\ket{\Psi_{P}}=0,
    \label{due-funf}
    \end{equation}
     where $H_{PP}=PHP$ etc.,
     \begin{equation}
      H'_{PP}=H_{PQ}(E-H_{QQ}+i\delta)^{-1}H_{QP}\,,
     \label{due-sech}
     \end{equation}
    and $\delta$ is a positive infinitesimal. The term $H'_{PP}$ in Eq. 
    (\ref{due-funf}) describes the Feshbach resonances. It represents an 
    effective non-local, retarded interaction in the P subspace due to 
    transitions to the Q subspace and back. The Hamiltonian $H_{PP}+H'_{PP}$ 
    gives the effective atom-atom interaction in the open-channel subspace, 
    the effects due to the existence of bound states being taken into account 
    through the term $H'_{PP}$.

    \subsection{Feshbach resonances in the 2D coupling}

     A Feshbach resonance results when true bound states belonging to the 
     closed-channel subspace match the energy of open channels: transient 
     transitions may then be possible during the collision process 
     \cite{moerdijk95}. Setting $H_{PP}=H_0 + U_1$ where $H_0$ includes the 
     kinetic energy of relative motion and the internal Hamiltonian, the 
     scattering matrix elements in the P subspace between plane-wave states 
     with relative momenta ${\bf k}$ and ${\bf k}'$ read
     \begin{equation}
     \bra{{\bf k}'}T\ket{{\bf k}}\simeq
     \bra{{\bf k}'}T_1\ket{{\bf k}}+\bra{\Omega_{1,+}{\bf k}'}H'_{PP}
     \ket{\Omega_{1,+}{\bf k}}\,.
     \label{due-intermedia}
     \end{equation}
     Here $T_1=U_1+U_1G_0T_1$ represents the scattering process if Q subspace
     is neglected and $\Omega_{1,+}=(1-G_0U_1)^{-1}$ is the wave operator
     due to the interaction potential $U_1$, with $G_0=(E-H_0)^{-1}$ 
     \cite{pethick}. In Eq. (\ref{due-intermedia}) the effect of closed 
     channels has been included at first order in $H'_{PP}$. In the limit of 
     zero relative velocity Eq. (\ref{due-intermedia}) leads to an expression 
     for the coupling strength,
     \begin{equation}
     g=\tilde g+\sum_{n}\frac{|\bra{\psi_{n}}
      H_{QP}\ket{\psi_{0}}|^2}{E_{th}-E_{n}}
     \label{olala}
     \end{equation}
     where the sum is over all the states $\ket{\psi_n}$ in Q subspace
    and we have set $\ket{\Omega_{1,+}{\bf k}}\simeq\ket{\Omega_{1,+}{\bf k}'}
    \equiv\ket{\psi_0}$. Here, $\tilde g$ is the coupling strength estimated 
    in the absence of closed channels and $E_{th}$ is the threshold energy for 
    the state $|\psi_0\rangle$ at vanishing kinetic energy. Finally, when the 
    threshold energy is close to the energy $E_{res}$ of one specific bound 
    state $\ket{\psi_{res}}$, this contribution will be dominant and the 
    contributions from all other states may be included in  $\tilde g$ through 
    an effective non-resonant scattering length $a_{nr}$. For a strictly 2D 
    condensate we get
    \begin{equation}
    g = \frac{4\pi\hbar^2/m}{\ln |4\hbar/\mu m a^2_{nr}|}+ 
    \frac{|\bra{\psi_{res}}H_{QP}\ket{\psi_{0}}|^2}{E_{th}-E_{res}} \,,
   \label{due-katz}
   \end{equation}
   where $\mu$ is the chemical potential of the condensate. 
   Equation (\ref{due-katz}) is our main result. The non-resonant first term 
   has been calculated from the two-body T-matrix for a binary collision 
   occurring at low momenta and energy in the presence of a condensate, which 
   fixes the energy of each colliding atom at the chemical potential 
   \cite{morgan,rajagopal}. The resonant second term can be tuned by 
   exploiting the dependence of the energy denominator on external parameters 
   and in particular on the strength of the applied magnetic field. 
   Explicitly, if the energy denominator vanishes at a value $B_0$ of the 
   field, then by expansion around this value we have
   \begin{equation}
   E_{th}-E_{res}\approx ({\frak{m}}_{res}-{\frak{m}}_{\alpha}
   -{\frak{m}}_{\beta})(B-B_{0})
   \label{due-ice}
   \end{equation}
   where ${\frak{m}}_{\alpha,\beta}=-{\partial\epsilon_{\alpha,\beta}}/
   {\partial B}$ are the magnetic moments of the two colliding atoms in the 
   open channel and ${\frak{m}}_{res}=-{\partial E_{res}}/{\partial B}$ is the 
   magnetic moment of the molecular bound state. Equation (\ref{due-katz}) 
   then yields
   \begin{equation}
   g = \frac{4\pi\hbar^2}{m} 
   \left( \frac{1}{\ln |4\hbar/\mu m a^2_{nr}|}+ 
   \frac{\Delta B}{B-B_{0}}\right)\,,
   \label{due-jinx}
   \end{equation}
   where the width parameter $\Delta B$ is defined as
   \begin{equation}
   \Delta B =\frac{m}{4\pi\hbar^2} 
    \frac{|\bra{\psi_{res}}H_{QP}\ket{\psi_{0}}|^2}
    {\frak{m}_{res}-\frak{m}_{\alpha}-\frak{m}_{\beta}}\,.
    \label{due-prat}
    \end{equation}
   Higher-order terms in $H'_{PP}$  will lead to a broadening of the 
   resonance. However, the width of resonant states close to the threshold 
   energy in the open channel is usually small because of the low density of 
   states, and Feshbach resonances are usually very sharp \cite{pethick}.
   
   In the next Section I will analyze the consequences of tuning 
   the coupling strength on the equilibrium state of a 2D condensate in a 
   harmonic trap.
 
   \subsection{Equilibrium properties as functions of the coupling}
   \label{sec2}

   We consider a Bose gas with a fixed number $N$ of particles in a 2D 
   harmonic potential $V_{ext}(r)=m\omega^2 r^2/2$. At zero temperature the 
   condensate wave function $\Psi(r)$ is described by the mean-field 
   Gross-Pitaevskii equation    
   \begin{equation}
   \left[-\frac{\hbar^2}{2m}\frac{1}{r}\frac{\partial}{\partial r}
   \left(r \frac{\partial}{\partial r}\right)+V_{ext}(r)+g n_{c}(r)\right]
   \Psi(r)=\mu \Psi(r) \,,
   \label{due-pasta}
   \end{equation}
   where $n_{c}(r)=|\Psi(r)|^2$ is the density of the condensate. We introduce 
   the dimensionless coordinate $x=r/a_{ho}$, where 
   $a_{ho}=(\hbar/m\omega)^{1/2}$ is the harmonic oscillator length, and the 
   dimensionless chemical potential $\mu'=2 \mu/\hbar\omega$. Thus 
   Eq. (\ref{due-pasta}) can be written in rescaled form as
   \begin{equation}
   \left[-\frac{d^2}{d x^2}- \frac{1}{x}\frac{d}{d x}\right]\Psi(x)
    + x^2\Psi(x) + 2\eta \left|\frac{\Psi(x)}{\Psi(0)}\right|^2\Psi(x)=
   \mu' \Psi(x) \,,
   \label{due-franca}
   \end{equation}
    where $\eta = g\,|\Psi(0)|^2/\hbar\omega$ is a dimensionless coupling 
    parameter given by the ratio between the mean-field interaction energy per 
    particle and the oscillator level spacing. The normalization condition on 
    the wave-function is
    \begin{equation}
     N = a_{ho}^2\int_{0}^{\infty}x\,dx \,|\Psi(x)|^2\,.
     \label{due-gatto}
     \end{equation}
     Equations (\ref{due-franca}) and (\ref{due-gatto}) need to be solved 
     self-consistently. 

     \subsection{Repulsive coupling ($\eta > 0$)}
     \label{code-bil}

\begin{figure}
     \centering{
      \epsfig{file=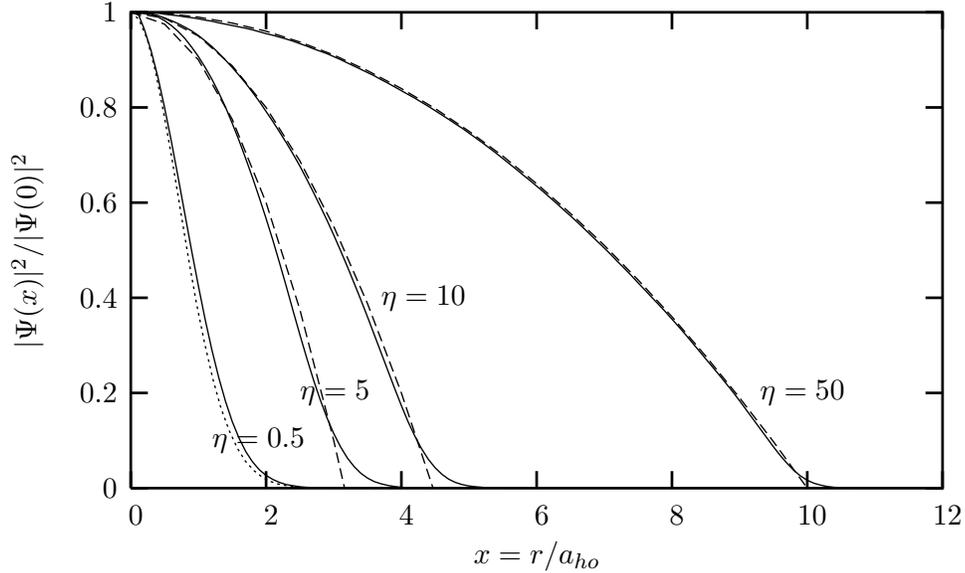,width=1.0\linewidth}}
      \caption{Scaled density profile $|\Psi(x)|^2/|\Psi(0)|^2$ as a function
       of the dimensionless radial distance $x=r/a_{ho}$ for various values of 
      the coupling parameter $\eta$. Dashed curves are Thomas-Fermi profiles 
      and the dotted curve is the free-gas profile.}
     \label{due-figfeshy}
     \end{figure}

     In the case of repulsive interactions Eq. (\ref{due-franca}) always 
     admits a solution: the condensate exists and is stable in the 2D 
     harmonic trap. Of course, far away from the resonance the dimensionless 
     coupling parameter $\eta$ is primarily determined by the non-resonant 
     scattering length. On approaching the resonance by increasing $B$ 
     towards $B_0$, the coupling parameter becomes very small for 
     $B\sim B_0-\Delta B/(\ln |4\hbar/\mu m a^2_{nr}|)$: the non-linear term 
     in Eq. (\ref{due-franca}) becomes small and the condensate wave function 
     is close to a Gaussian profile. On further increasing $B$ the gas enters 
     the strong-coupling regime and in the limit $\eta \gg 1$ the kinetic 
     energy term in Eq. (\ref{due-franca}) and the discrete structure of the 
     trap levels become irrelevant. The profile then approaches a Thomas-Fermi 
     form \cite{goldman,huse}. The shape of the condensate density profile, 
     obtained by the numerical solution of Eq. (\ref{due-franca}) at 
     several values of the coupling parameter $\eta$, is depicted in 
     Fig. \ref{due-figfeshy}. The Figure illustrates comparisons with the 
     free-gas and Thomas-Fermi profiles, according to the discussion given 
     just above.

     \subsection{Attractive coupling ($\eta < 0$)}

      On going through the resonance the Thomas-Fermi-like density profile 
      will suddenly shrink and collapse. Indeed, in the strong-coupling 
      attractive regime the kinetic energy does not suffice to compensate for 
      the negative mean-field energy term and Eq. (\ref{due-pasta}) does not 
      admit physical solutions. A condensate with a limited number of 
      particles can only exist in the weakly attractive regime. An upper bound 
      for the critical number $N_c$ of bosons in a condensate with weakly 
      attractive interactions can be estimated by comparing the kinetic energy 
      and the mean-field energy at the centre of the trap. This leads to the 
      inequality
      \begin{equation}
      1+2\eta\ge0
      \label{due-samba}
      \end{equation}
      for condensate stability, where in the calculation of the kinetic energy 
      we have used the free-gas wave function. Using the expression of $\eta$ 
      we find from Eq. (\ref{due-samba})
      \begin{equation}
      N_c\simeq\left|\frac{1}{\ln |4\hbar/\mu m a^2_{nr}|}+ 
      \frac{\Delta B}{B-B_{0}}\right|^{-1}.
      \label{due-singor}
      \end{equation}
     At variance from the trapped 3D gas \cite{antonius}, the critical number 
     $N_c$ does not depend on the harmonic oscillator length, the dependence 
     being cancelled by the fact that in 2D the mean-field energy scales as 
     $\approx a_{ho}^{-2}$ like the kinetic energy. Let us remark that the 
     condensate in the case of attractive coupling can be metastable even if 
     the number of condensed bosons is lower than $N_c$. A three-body loss 
     term should be included in Eq. (\ref{due-pasta}) in order to calculate 
     the decay time and to describe the time evolution of such a condensate, 
     as was done for the 3D Bose gas \cite{savage,saito,sadhan} and very 
     recently for a 3D boson-fermion mixture \cite{adikari}. An evaluation of 
     three-body losses in the 2D Bose gas and of the time evolution of a 2D 
     condensate in the weakly attractive coupling regime may be considered in 
     the future.

%%%%%%%%%%%%%%%%%%%%%%%%%%%%%%%%%%%%%%%%%%%%%%%%%%%%%%%%%%%%
% \clearpage{\pagestyle{empty}\cleardoublepage}
 \chapter[Rotating 2D trapped Bose gas]{Rotating 2D trapped Bose gas} 
 \label{fcs_shuttle_chapter}
\section{Single vortex in a rotating 2D Bose gas}
\label{single vortex}

Understanding the behaviour of vortices in the 2D regime is important, since 
they reflect the superfluid nature of the condensate 
\cite{lundh,castin,dalfovo} and are expected to play a role in the 
transition from the superfluid to the normal state \cite{kosterlitz}.
An experimental method for the creation of quantized vortices in a trapped BEC 
has  made use of an "optical spoon" \cite{madison1,rosenbusch}, whereby the 
condensate in an elongated trap is set into rotation by stirring with laser 
beams. This experiment on a confined boson gas is conceptually the analogue of 
the rotating bucket experiment on bulk superfluid Helium, with the difference 
that the thermal cloud is inhomogeneously distributed and lies mostly outside 
the condensate cloud. A quantized vortex first appears in the condensate at a 
critical angular frequency of stirring which corresponds to an instability of 
the vortex-free state. A lower bound for the critical frequency can be 
assessed from the energy of a vortex, defined as the difference in the 
internal energy of the gas with and without a vortex. Observation of a vortex 
in a trapped gas is difficult since the vortex core is small in comparison to 
the size of the boson cloud, but the size of the core increases during free 
expansion and indeed vortices were first observed experimentally by releasing 
the trap and allowing ballistic expansion of the cloud \cite{madison1}.

\subsection{GP model that accomodates a single vortex}

The BEC is subject to an anisotropic harmonic confinement characterized 
by the radial trap frequency $\omega_\perp$  and by the axial frequency 
$\lambda\omega_\perp$  with $\lambda\gg 1$. Motions along the $z$ 
direction are suppressed and the condensate wave function is determined 
by a 2D equation of motion in the $\{x,y\}$ plane.

The order parameter for a 2D condensate accommodating a quantized vortex 
state of angular momentum $\hbar\kappa$ per particle is written as 
$\Psi({\bf r})=\psi(r)\exp(i\kappa\phi)$, with $\phi$ the azimuthal 
angle. The wave function $\Psi(r)$ then obeys the nonlinear Schr\"odinger 
equation (NLSE)
\begin{equation}
\left[-{\hbar^2\over 2m}
\nabla^2+{\hbar^2\kappa^2\over 2m r^2}+{1\over 2}m\omega_\perp^2r^2+
g_2n_c(r)+2g_1n_{nc}(r)\right]\Psi(r)=\mu\Psi(r)\, 
\label{tre-GP2}
\end{equation}
similar to Eq. (\ref{one-GP}) but in addition it includes the contribution 
$2g_1n_{nc}(r)$ of the thermal clouds. The 2D coupling parameters $g_j$, with 
$j = 2$ for condensate-condensate repulsions and $j = 1$ for 
condensate-noncondensate repulsions, are given by
\begin{equation}
    g_{j}^{2B}= \frac{4\pi\hbar^2/m}{\ln|4\hbar^2/(jm\mu a^{2})|}\,.
   \label{tre-2bcoup}
   \end{equation}
In Eq. (\ref{tre-2bcoup}) above we have omitted a term due to thermal 
excitations \cite{rajagopal}, which is negligible in the temperature range of 
present interest ($T\le 0.5 T_c$, with $T_c$ the critical temperature).

In this model the atoms in the thermal cloud are not put directly 
into rotation, but feel the rotating condensate through the mean-field 
interactions. In the Hartree-Fock approximation \cite{minguzzi97} 
the thermal cloud is treated as an ideal gas subject to the effective 
potential $V_{eff}(r)$ written as in Eq. (\ref{two-veff}). The density 
distribution of the thermal cloud is then given by Eq. (\ref{two-Tden}) in 
which the momentum cut-off $p_0$ in Eq. (\ref{two-Tden}) is equivalent to 
adding the term $2g_1n_{nc}$ to the effective potential in 
Eq. (\ref{two-veff}).

We solve self-consistently the coupled Eqs. (\ref{tre-GP2}) and 
(\ref{two-Tden}) together with the condition that the areal integral of 
$n_{c}(r)$+$n_{nc}(r)$ is equal to the total number $N$ of particles. The 
differential equation (\ref{tre-GP2}) is solved iteratively by discretization, 
using a two-step Crank-Nicholson scheme \cite{press}. In 
Sec. (\ref{tre-numres}) I will also compare the results with those obtained 
in the Thomas-Fermi approximation by dropping the radial kinetic energy term 
in Eq. (\ref{tre-GP2}).

\subsection{The energy of a vortex}

The critical angular velocity measured in experiments where vortex nucleation 
occurs from a dynamical instability \cite{madison1} depends strongly on the 
shape of the perturbation. However, a lower bound for the angular velocity 
required to produce a single-vortex state can be estimated once the energies 
of the states with and without the vortex are known. Since the angular 
momentum per particle is $\hbar\kappa$, the critical (thermodynamic) angular 
velocity is given by 
\begin{equation}
\Omega_c=\frac{E_\kappa-E_{\kappa=0}}{N\hbar\kappa}\,.
\label{tre-critfreq}
\end{equation}
This expression follows by equating the energy of the vortex state in the 
rotating frame, that is $E_\kappa-\Omega_cL_z$, to the energy $E_{\kappa=0}$ 
of the vortex-free state.

In the noninteracting case at zero temperature, the energy difference per 
particle is simply $\hbar\kappa\omega_\perp$, so that $\Omega_c$ is just the 
trap frequency in the $\{x,y\}$ plane. For the interacting gas at zero 
temperature in the Thomas-Fermi approximation, Eq. (\ref{tre-critfreq}) 
reduces to the expression \cite{baym96,lundh}
\begin{equation}
\Omega_c^{TF}(0)=\frac{2\hbar}{m R^2}\ln\left(\frac{0.888\,R}{\xi}\right)\,.
\label{tre-formula}
\end{equation}
Here $R=(2\mu/m\omega_\perp^2)^{1/2}$ and $\xi=R \,\hbar\omega_\perp/2\mu$ 
are the Thomas-Fermi radius and the healing length respectively.

In the general case of an interacting gas at finite temperature, we have 
to evaluate numerically the total energy as the sum of four 
terms \cite{giorgini},
\begin{equation}
E_\kappa=E_{\rm kin,c}+E_{\rm trap}+E_{\rm int}+E_{\rm kin,T}\,.
\label{tre-energy}
\end{equation}
These terms are the kinetic energy of the condensate
\begin{equation}
E_{\rm kin,c}=\int d^2r \,\psi^*(r)\left(-{\hbar^2\over 2m}
\nabla^2+{\hbar^2\kappa^2\over 2m r^2}\right)\psi(r)\,,
\label{tre-kinetic}
\end{equation}
the energy of confinement
\begin{equation}
E_{\rm trap}=\frac{1}{2}m\omega_\perp^2
\int d^2r\,r^2[n_c(r)+n_{nc}(r)]\,,
\label{tre-trap}
\end{equation}
the interaction energy
\begin{equation}
E_{\rm int}=\frac{1}{2}\int d^2r[g_2n^2_c(r)+4g_1n_c(r)n_{nc}(r)
+2g_1n_{nc}^2(r)]\,,
\label{tre-interac}
\end{equation}
and the kinetic energy of the thermal cloud
\begin{equation}
E_{\rm kin,T}=\int d^2r \int \frac{dp}{2\pi\hbar^2} \frac{p^3}{2m}
\left\{\exp[\beta(p^2/2m+V_{eff}(r)-\mu)]-1\right\}^{-1}.
\label{tre-thermal}
\end{equation}

In Sec. (\ref{tre-numres}) I compare the angular frequency obtained by the 
calculation of the total energy of the gas with and without a vortex with 
those obtained from a Thomas-Fermi calculation and from extrapolating 
Eq. (\ref{tre-formula}) at finite temperature through the temperature 
dependence of the chemical potential.

\subsection{Numerical result for single vortex }
\label{tre-numres}
\begin{figure}
\centering{\epsfig{file=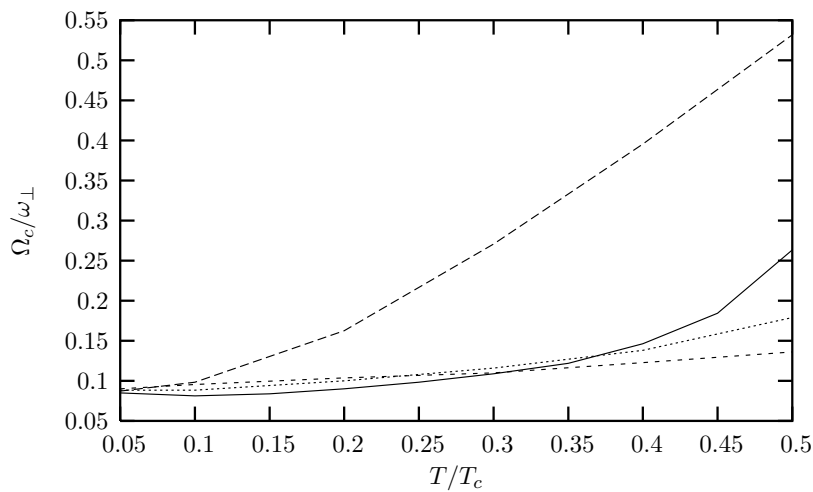,width=0.8\linewidth}}
\caption{The energy of a vortex, expressed as the thermodynamic critical 
frequency $\Omega_c$ in units of 
the radial trap frequency $\omega_\perp$, as a function of $T/T_c$. 
The results from the full calculation using the NLSE (full line) are 
compared with those obtained from Eq. (\ref{tre-formula}) with the 
corresponding values of $\mu(T)$ (dotted line) and with the Thomas-Fermi 
values of $\mu(T)$ (short-dashed line). The long-dashed line shows the 
results obtained in a calculation using the Thomas-Fermi theory.}
\label{tre-figbmw}
\end{figure}

For a numerical illustration we have taken $\kappa = 1$ and chosen values 
values of particle number, the radial trap frequency, and the scattering 
length as appropriate for $^{23}$Na atoms in the experiment of 
G\"orlitz {\it et al.}\cite{gorlitz}
($N =5\times 10^5,\,\omega_{\perp}=188.4\,$Hz, $a=2.8$ nm). I implicitly 
assume, however, that the trap has been axially squeezed to reach the strictly 
2D scattering regime.
\begin{figure}
\centering{
\epsfig{file=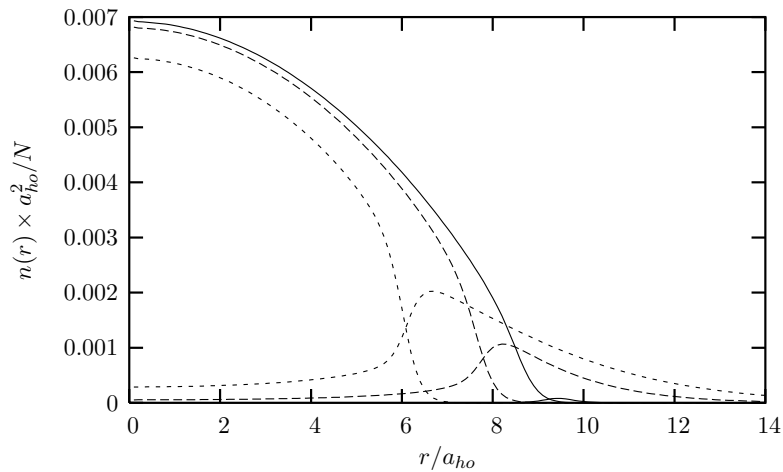,width=0.8\linewidth}}
\caption{Density profiles $n(r)$ for the condensate and the thermal cloud 
(in units of $a_{ho}^2/N$, with $a_{ho}=\sqrt{\hbar/m\omega_\perp}$) 
{\it versus} radial distance $r$ (in units of $a_{ho}$) 
at temperature $T/T_c= 0.05$ (full line), 0.25 (long-dashed line), 
and 0.50 (short-dashed line).}
\label{tre-figfiat}
\end{figure}
The density profiles obtained from the solution of Eqs. (\ref{tre-GP2})
for the gas at three different values of the temperature (in units of the 
critical temperature $T_c=(\sqrt{6N}/\pi k_B)\hbar\omega_\perp$ of the ideal 
Bose gas) are shown in Figs. \ref{tre-figfiat} and \ref{tre-figopel}. In the 
absence of a vortex (Fig. \ref{tre-figfiat}), the main point to notice is that 
the growth of the thermal cloud exerts an increasing repulsion on the outer 
parts of the condensate, constricting it towards the central region of the 
trap. This effect is seen only when the condensate is treated by the NLSE and 
persists in the presence of a vortex (Fig. \ref{tre-figopel}), but as we 
shall see below, is missed in the Thomas-Fermi approximation where the radial 
kinetic energy of the condensate is neglected. In addition, the thermal cloud 
penetrates the core of the vortex and enhances the expulsion of the condensate 
from the core region (inset of Fig. \ref{tre-figopel}), in a manner which 
again is governed in its details by the radial kinetic energy term in the NLSE.
\begin{figure}
\centering{
\epsfig{file=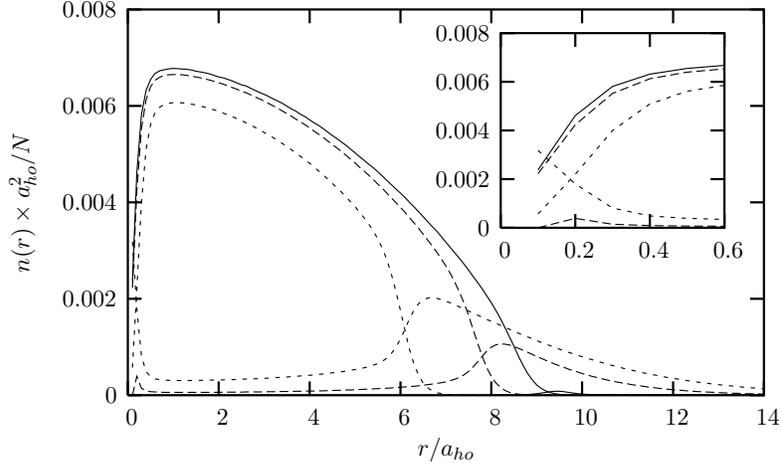,width=0.8\linewidth}}
\caption{Density profiles for the condensate and for the thermal cloud 
in the presence of a vortex. Units and symbols are as in 
Fig. \ref{tre-figfiat}. The inset shows an enlarged view of the 
profiles near the center of the trap.}
\label{tre-figopel}
\end{figure}
The profiles obtained for a BEC containing a vortex at $T = 0.5\,T_c$  
by the full numerical calculation using the NLSE are compared in 
Fig. \ref{tre-figmazda} with those obtained in the Thomas-Fermi 
approximation. The constriction of the condensate in its outer parts 
and its expulsion from the core region by the thermal cloud are clearly 
underestimated in the Thomas-Fermi theory.
\begin{figure}
\centering{\epsfig{file=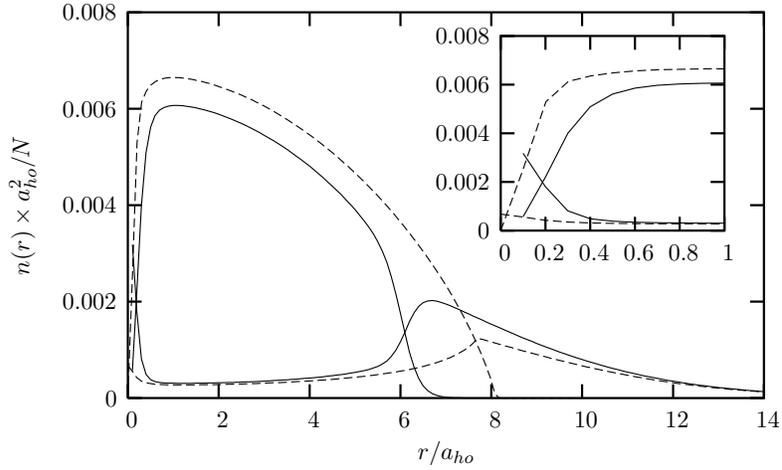,width=0.8\linewidth}}
\caption{Density profiles for a BEC containing a vortex at $T/T_c= 0.50$ 
in the full calculation using the NLSE (full lines) and in the 
Thomas-Fermi approximation (dashed lines). The units 
are as in Fig. \ref{tre-figfiat}. The inset shows an enlarged view of the 
profiles near the center of the trap.}
\label{tre-figmazda}
\end{figure}
Finally, Fig. \ref{tre-figbmw} reports our results for the energy of the 
vortex as a function of $T/T_c$. The inaccuracies arising in the density 
profiles from the Thomas-Fermi treatment of the interplay between the 
condensate and the thermal cloud clearly lead to large errors in the 
estimation of $\Omega_c(T)$.
On the other hand the simple expression given in Eq. (\ref{tre-formula}), with 
the two alternatives of using in it the chemical potential $\mu(T)$  from 
the Thomas-Fermi approximation or from the full calculation using the NLSE, 
gives a reasonable account of the vortex energy up to $T\simeq0.5\,T_c$.

%\section{Dynamics of vortices for a 2D Bose fluid}
%\label{many vortex}

\section{Density Functional Theory for many-vortex problem}

\subsection{Introduction}

Nelson et al. \cite{nelson1} have shown that the statistical mechanics of the 
flux-line lattice (FLL) of high-$T_{c}$ superconductors can be studied through 
an appropriate mapping onto the 2D Yukawa boson system interacting with
potential $V_{Y}(r)=\epsilon K_{0}(r/\sigma)$ where $\epsilon$ is an energy 
scale and $\sigma$ is a length scale having the meaning of a screening length. 
The function $K_{0}(x) $ being the modified Bessel function and the 
coefficient $\epsilon$ attached to it is the coupling strength that scales the 
energy of interaction.

The Hamiltonian of $N$ bosons describing this system reads
\begin{equation}
H= -\sum_{i}\Lambda^2{\bf\nabla}_{i}^2 +\sum_{i<j}K_{0}(r_{ij}) \,.
\label{Ham1}
\end{equation}
Here $i$, $j$ are particle indices and the de Boer dimensionless 
parameter is defined by $\Lambda^{2}=\hbar^2/2m\sigma^2\epsilon$. Meanwhile
the reduced density of the system is denoted by $\rho=N/A\sigma^{2}$, where 
$A$ is the surface area. These two dimensionless parameter characterize the 
system. 

On the other hand, the Hamiltonian for a system of $N$ vortices mapped as a 
system of $N$ boson interacting with Yukawa potential is written as
\begin{equation}
H= -\sum_{i}\frac{(k_{B}T)^2}{2\tilde{\epsilon}}{\bf\nabla}_{i}^2 
+\sum_{i<j}\frac{\phi_{0}^2}{8\pi^2\lambda^2}K_{0}(r_{ij}/\lambda)\,.
\label{Ham2}
\end{equation}
The system described by the Hamiltonian (\ref{Ham2}) is equal to one 
described by Hamiltonian (\ref{Ham1}) if we set $\sigma=\lambda$ and 
$\Lambda^2=(2\pi k_{B}T)^2)/2\tilde{\epsilon}\phi_{0}^2$. \par

\begin{figure}
\centering{
\epsfig{file=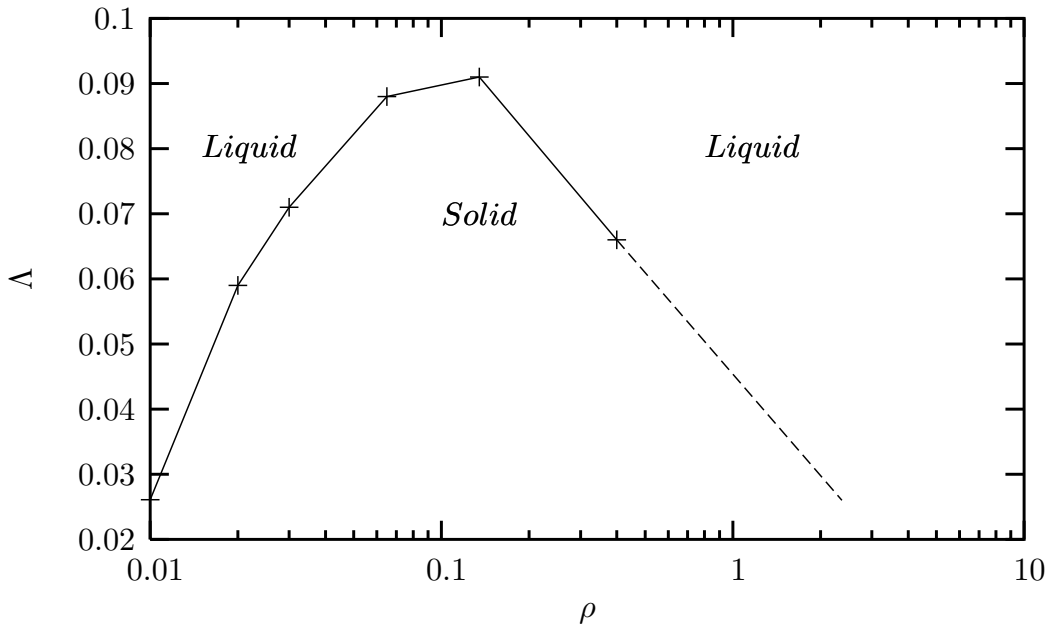,width=1.0\linewidth}}
\caption{The phase diagram of Yukawa bosons reproduced from Magro and 
Ceperley \cite{Magro93a}. The pluses are transition points computed with DMC. 
The dashed lines at high densities is the scaling law: 
$\Lambda\sim 0.04/\sqrt{\rho}$. }  
\label{fig_phase}
\end{figure}

Following this idea, Magro and Ceperley \cite{Magro93a,Magro94a} have 
performed a Diffusion Monte Carlo (DMC) and Variational Monte Carlo (VMC) 
numerical calculation to calibrate the ground state vortex properties for a 
homogeneous 2D Yukawa boson system. Of particular interest to us is the phase 
diagram obtained for the data of ($\rho$, $\Lambda$) at transition points 
displayed in Fig (\ref{fig_phase}).

They observed that for a particular region $\Lambda > 0.09$ the system is 
dominated by kinetic energy and does not crystallize. Below this threshold 
however a peculiar behaviour of reentrant liquid has been observed. Which 
means that system is in the liquid phase at very low density as well at high 
density. The crystal melts on compression and expansion. Nevertheless this 
reentrant liquid behaviour has not been fully understood.  

Inspired by their work, I will investigate the ground state property of a 
system of $N$ vortices mapped as $N$ bosons interacting $via$ 2D Yukawa 
potential $V(x)=\epsilon\,K_{0}(x)$ confined in a 2D harmonic planar geometry 
$\frac{1}{2}m\omega_{\perp}^2 {\bf r}^2$. As a matter of fact, the Yukawa 
potential has an interesting combination of short range with soft core. For 
small $x$, $K_{0}(x)$ diverges like $-\ln(x)$ while for large $x$ it decays 
as $\exp(-x)/\sqrt{x}$. The logaritmic term represents the inter-vortex 
repulsion and the exponential decay signifies that vortices far apart do 
not feel each other.

\subsection{Density Functional Theory }

With the Yukawa potential we cannot use the similar approach used for the 
contact delta potential $V_{0}\delta({\bf r}-{\bf r'})$  in deriving a 
dynamical equation of motion as described in chapter 1. Here we use 
the Density functional theory (DFT) which is originally based on the notion 
that for a many-electron system there is a one-to-one mapping between the 
external potential and the electron density: $v_{ext}({\bf r})$ 
$\leftrightarrow$ $\rho({\bf r})$. In other words, the density is uniquely 
determined given a potential, and $vice\,\,versa$. All properties are 
therefore a functional of the density, because the density determines the 
potential, which determines the Hamiltonian, which determines the energy and 
the wave function .

Following this train of reasoning, the inhomogeneous dilute system of $N$ 
interacting bosons can be described within the second quantization 
language as
\begin{eqnarray}
\hat{H} &=& \hat{H_{0}}+\int\,d{\bf r} \psi^\dagger({\bf r})V_{ext}({\bf r})
\psi({\bf r})\nonumber \\
&+&\frac{1}{2}\int\,d{\bf r}\int\,d{\bf r'}
\psi^\dagger({\bf r})\psi^\dagger({\bf r'})V(|{\bf r}-{\bf r'}|)
\psi({\bf r'})\psi({\bf r})\nonumber \\
&=& \hat{H_{0}}+\hat{V}_{ext}+\hat{V}_{int}
\label{vor-a}
\end{eqnarray}    
where $\hat{H_{0}}=\int\,d{\bf r} \psi^\dagger({\bf r})
\left[-\frac{\hbar^2}{2m}\nabla^2-\mu \right ]\psi({\bf r})$
 and $V(|{\bf r}-{\bf r'}|)$ is the inter-atomic interaction potential. The 
annihilation and creation field operators are denoted by 
$\psi^\dagger({\bf r})$ and $\psi({\bf r'})$ respectively and obey 
Bose-Einstein commutation relations:
\begin{equation}
[\psi({\bf r}),\psi^{\dagger}({\bf r'})]=\delta({\bf r}-{\bf r'})\,\,\,\,\,\,
\,\,[\psi({\bf r}),\psi({\bf r'})]=
[\psi^{\dagger}({\bf r}),\psi^{\dagger}({\bf r'})]=0\,.
\label{vor-b}
\end{equation}

Let us denote the ground state of the system as $\ket{g}$ so the ground state
energy is defined as $E_{0}=\bra{g}\hat{H}\ket{g}$ and the density as 
$n({\bf r})=\bra{g}\psi^{\dagger}\psi\ket{g}$. The Hohenberg-Kohn (HK) theorem 
\cite{HK64} guarantees that there exists a unique functional of the density, 
\begin{equation}
F[n({\bf r})]=\bra{g}\hat{H_{0}}+\hat{V}_{int}\ket{g}\,,
\label{Fenergy}
\end{equation}
irrespective of the external potential. The theorem was originally proved for 
fermions but its generalization also covers bosons. Following HK, we can write 
the total energy functional of the system as following,
\begin{equation}
E[n({\bf r})]=F[n({\bf r})]+ \int d{\bf r}V_{ext}({\bf r})n({\bf r})
\label{Tenergy}
\end{equation}
Determination of the ground state energy follows by imposing the stationary 
conditions
\begin{equation}
\frac{\delta E[n({\bf r})]}{\delta n({\bf r})}=0
\label{vari-Primo}
\end{equation} 
In general, the ground state cannot be determined exactly, but one has to 
resort to the Kohn-Sham procedure \cite{KS65} to introduce an accurate 
approximation. The idea is to map the interacting system of interest to an 
auxiliary system. 
In order to describe the auxiliary system, let us first decompose the Bose 
quantum field operators $\psi({\bf r})$ as a sum of spatially varying 
condensate $\Phi({\bf r})$ and a fluctuation field operator $\tilde{\psi}$ :
\begin{equation}
 \psi({\bf r})= \Phi({\bf r})+ \tilde{\psi}({\bf r}) \,\,\,\,\,
, \psi^{\dagger}({\bf r})= \Phi^{*}({\bf r})+ \tilde{\psi}^{\dagger}({\bf r})
\label{vor-d}
\end{equation}
The wave function $\Phi({\bf r})$ is also known as the order parameter and 
is defined as the statistical average of the particle field operator
$\Phi({\bf r})=<\psi>$. On the other hand, the condensate density is defined 
in term of order parameter as $n_{c}({\bf r})=|\Phi({\bf r})|^2$. In the 
well-known Bogoliubov approximation \cite{bogoliubov} for a dilute weakly 
interacting gas, in which almost all the 
atoms are Bose condensed, one keeps only the term up to quadratic in the 
non-condensate field operators $\tilde{\psi}$ and $\tilde{\psi}^{\dagger}$.
When the fluctuation field is comparably small to the spatial order parameter
(in an average sense, $\tilde{\psi}\ll\Phi$) one can comfortably utilise a 
Mean-field approach. Using the decomposition in Eq.(\ref{vor-d}) above and 
applying the Bogoliubov approximation, one can replace the product of field 
operators 
$\psi^{\dagger}({\bf r})\psi^{\dagger}({\bf r'})\psi({\bf r'})\psi({\bf r})$
in the interaction energy term of Eq.(\ref{vor-a}) by
\begin{eqnarray}
\psi^{\dagger}({\bf r})\psi^{\dagger}({\bf r'})\psi({\bf r'})\psi({\bf r})
&=& |\Phi({\bf r})|^2|\Phi({\bf r'})|^2 +2|\Phi({\bf r})|^2 
\left[ \Phi({\bf r'})\tilde{\psi}^{\dagger}({\bf r'})+\Phi^{*}({\bf r'})
\tilde{\psi}({\bf r'})\right]
\nonumber \\
&+& |\Phi({\bf r'})|^2\tilde{\psi}^{\dagger}({\bf r})
\tilde{\psi}({\bf r})+ \Phi^{*}({\bf r'})\Phi({\bf r})
\tilde{\psi}^{\dagger}({\bf r})\tilde{\psi}^{\dagger}({\bf r'})
\nonumber \\
&+&\frac{1}{2}\Phi^{*}({\bf r'})\Phi^{*}({\bf r})\tilde{\psi}({\bf r})
\tilde{\psi}({\bf r'})+\frac{1}{2}\Phi({\bf r'})\Phi({\bf r})
\tilde{\psi}^{\dagger}({\bf r})\tilde{\psi}^{\dagger}({\bf r'}) \,.  
\label{IntTot}
\end{eqnarray}
So now we are in a position to define the Hamiltonian of the auxiliary 
system that includes the inter-particle potential defined above as
\begin{equation}
\hat{H}^{s}= \int d{\bf r}\, \psi^\dagger({\bf r})\left[-\frac{\hbar^2}{2m}
\nabla^2-\mu \right ]\psi({\bf r}) + \hat{V_{s}}+
\int d{\bf r}\, \psi^\dagger({\bf r})V_{ext}^{s}({\bf r})\psi({\bf r})\,.
\label{AuxHamil}
\end{equation}
where our definition of auxiliary interparticle potential reads
\begin{eqnarray}
\hat{V_{s}}&=& \int d{\bf r}\,\int d{\bf r'}\,V(|{\bf r}-{\bf r'}|)
|\Phi({\bf r})|^2 \left[ \Phi({\bf r'})\tilde{\psi}^{\dagger}({\bf r'})
+\Phi^{*}({\bf r'})\tilde{\psi}({\bf r'})\right]\nonumber \\
&+&\frac{1}{2}\int d{\bf r}\,\int d{\bf r'}\,V(|{\bf r}-{\bf r'}|)
 [2\Phi^{*}({\bf r'})\Phi({\bf r})\tilde{\psi}^{\dagger}({\bf r})
\tilde{\psi}({\bf r'})\nonumber \\
&+&\Phi({\bf r'})\Phi({\bf r})\tilde{\psi}^{\dagger}({\bf r})
\tilde{\psi}^{\dagger}({\bf r'})+\Phi^{*}({\bf r'})\Phi^{*}({\bf r})
\tilde{\psi}({\bf r})\tilde{\psi}({\bf r'})] \,.  
\label{AuxInt}
\end{eqnarray}
Thus for the ground state $\ket{g_{s}}$ we can define the unique ground state 
energy functional of the auxiliary system as
\begin{equation}
F^{s}[n_{s}({\bf r})]=\bra{g_{s}}\int d{\bf r} \psi^\dagger({\bf r})
\left[-\frac{\hbar^2}{2m}\nabla^2-\mu \right ]\psi({\bf r})
\ket{g_{s}} + \bra{g_{s}}\hat{V_{s}}\ket{g_{s}}\,,
\label{AuxFunc}
\end{equation}
which does not depend on the auxiliary external potential $V_{ext}^{s}$. 
Hence, the total energy functional of the auxiliary system can be written as
\begin{equation}
E^{s}[n_{s}({\bf r})]= F^{s}[n_{s}({\bf r})]+ \int d{\bf r}\, 
n({\bf r})V_{ext}^{s}({\bf r})
\label{AuxEnergy}
\end{equation}
in which $E^{s}[n_{s}({\bf r})]$ can be approximated to $E[n({\bf r})]$ or 
in other word the density of the auxiliary  system $n_{s}({\bf r})$ is 
identical to the real system $n({\bf r})$ by choosing a proper choice of 
auxiliary external potential $V_{ext}^{s}$. \par
Since the non-interacting part of Eq. (\ref{AuxFunc}) is equal to the 
non-interacting part of Eq. (\ref{Fenergy}) and comparing terms in 
Eq.(\ref{AuxInt}) with Eq. (\ref{IntTot}) we obtain the following functional 
relation:
\begin{equation}
F[n({\bf r})]= F^{s}[n({\bf r})]+ \int d{\bf r}\,\int d{\bf r}\,
V({\bf r}-{\bf r'})n({\bf r})n({\bf r'})+F_{xc}[n({\bf r})]\,.
\label{Frelation}
\end{equation}
The second term $\hat{V}_{H}$ is called the Hartree-energy defined as  
\begin{equation}
\hat{V}_{H}=\frac{1}{2}\int d{\bf r}\,\int d{\bf r'}\,V(|{\bf r}-{\bf r'}|)
 n({\bf r})n({\bf r'})
\label{Hartree}
\end{equation}
in which the total density $ n({\bf r})$ includes a non-condensate local 
density defined as
\begin{equation}
\tilde{n}({\bf r})\equiv <\tilde{\psi}^{\dagger}({\bf r})
\tilde{\psi}({\bf r})>= n({\bf r})-|\Phi({\bf r})|^2 \,.
\label{noncon}
\end{equation}
The last term in Eq. (\ref{Frelation}) represents the exchange-correlation 
energy $E_{xc}[n({\bf r})]$ and includes all the contributions to the 
interaction energy beyond mean field. Calculating the variational derivatives 
in Eq. (\ref{vari-Primo}) using Eq.(\ref{Frelation}), one finds
\begin{equation}
\frac{\delta F^{s}[n({\bf r})]}{\delta n({\bf r})}+V_{H}({\bf r})+
\frac{\delta F_{xc}[n({\bf r})]}{\delta n({\bf r})}+V_{ext}({\bf r})=0
\label{vari-ultimo}
\end{equation}
where the Hartree field read 
\begin{equation}
V_{H}({\bf r})=\int d{\bf r'}\, V(|{\bf r}-{\bf r'}|)n({\bf r'})\,.
\label{Hart2}
\end{equation}
Performing a similar variational calculation on Eq. (\ref{AuxEnergy}) we 
deduce that the density of the auxiliary system is identical to the 
actual system if 
\begin{equation}
V_{ext}^{s}({\bf r})=V_{ext}({\bf r})+ V_{H}({\bf r})+ 
\frac{\delta F_{xc}[n({\bf r})]}{\delta n({\bf r})}\,.
\label{trapRel}
\end{equation}

\subsection{Interacting auxiliary system : Bogoliubov approach}

In general, if an interacting auxiliary system is chosen, one needs to do
further analysis by using a Bogoliubov approach. We rewrite the auxiliary 
Hamiltonian Eq. (\ref{AuxHamil}) using the decomposition  in 
Eq. (\ref{vor-d}) by
\begin{eqnarray}
\hat{H}^{s}&=& \int\,d{\bf r} \Phi^{*}({\bf r})L\Phi({\bf r})+ \int\,d{\bf r} 
\tilde{\psi}^{\dagger}({\bf r})L\Phi({\bf r})+ \int\,d{\bf r} \Phi^{*}({\bf r})
L\tilde{\psi}({\bf r}) \nonumber \\
&+& \int\,d{\bf r} \tilde{\psi}^{\dagger}({\bf r})L\tilde{\psi}({\bf r})
+\frac{1}{2}\int\,d{\bf r}\int\,d{\bf r'}V_{Y}(|{\bf r}-{\bf r'}|)
 [2\Phi^{*}({\bf r'})\Phi({\bf r})\tilde{\psi}^{\dagger}({\bf r})
\tilde{\psi}({\bf r'})\nonumber \\
&+&\Phi({\bf r'})\Phi({\bf r})\tilde{\psi}^{\dagger}({\bf r})
\tilde{\psi}^{\dagger}({\bf r'})+\Phi^{*}({\bf r'})\Phi^{*}({\bf r})
\tilde{\psi}({\bf r})\tilde{\psi}({\bf r'})]\,
\label{newAuxHamil}
\end{eqnarray} 
where the operator $\hat{L}$ is defined by
\begin{equation}
\hat{L}\equiv \left( -\frac{\nabla^2}{2m}+V_{ext}^{s}-\mu \right) \,.
\label{operator}
\end{equation}
In order to diagonalize Eq. (\ref{newAuxHamil}), we first eliminate the terms 
linear in $\tilde{\psi}$ and $\tilde{\psi}^{\dagger}$ by requiring that 
$\Phi({\bf r})$ satisfy the equation 
\begin{equation}
\hat{L}\Phi({\bf r})=0 \,.
\label{LifePrimo}
\end{equation}
Hence our auxiliary Hamiltonian Eq. (\ref{newAuxHamil})reduces to 
\begin{eqnarray}
\hat{H}^{s}&=& \int\,d{\bf r} \tilde{\psi}^{\dagger}({\bf r})L
\tilde{\psi}({\bf r}) +\int\,d{\bf r}\int\,d{\bf r'}V(|{\bf r}-{\bf r'}|)
\Phi^{*}({\bf r'})\Phi({\bf r})\tilde{\psi}^{\dagger}({\bf r})
\tilde{\psi}({\bf r'})\nonumber \\
&+&\frac{1}{2}\int\,d{\bf r}\int\,d{\bf r'}V(|{\bf r}-{\bf r'}|)
[\Phi({\bf r'})\Phi({\bf r})\tilde{\psi}^{\dagger}({\bf r})
\tilde{\psi}^{\dagger}({\bf r'})\nonumber \\
&+&\Phi^{*}({\bf r'})\Phi^{*}({\bf r})
\tilde{\psi}({\bf r})\tilde{\psi}({\bf r'})]\,.
\label{newAuxHamil2}
\end{eqnarray}
The quadratic expression given by this equation can be diagonalized by the 
usual Bogoliubov transformation \cite{bogoliubov}
\begin{eqnarray}
\tilde{\psi}({\bf r})=\sum_{i}[u_{j}({\bf r})\alpha_{j}-v_{j}^{*}({\bf r})
\alpha_{j}^{\dagger}]\nonumber \\
\tilde{\psi}^{\dagger}({\bf r})=\sum_{i}[u_{j}^{*}({\bf r})\alpha_{j}^{\dagger}
-v_{j}({\bf r})\alpha_{j}]\,,
\label{bogo}
\end{eqnarray} 
where the quasiparticle operators $\alpha_{i}$ and $\alpha_{i}^{\dagger}$ 
satisfy Bose commutation relations. The values of $u_{i}$ and $v_{i}$ are 
obtained by solving the following coupled Bogoliubov equations:
\begin{eqnarray}
 \hat{L}u_{i} &+&\int d{\bf r'}\,V(|{\bf r}-{\bf r'}|)\Phi^{*}({\bf r'})
\Phi({\bf r})u_{i}({\bf r'})\nonumber \\
&+&\int d{\bf r'}\,V(|{\bf r}-{\bf r'}|)
\Phi({\bf r'})\Phi({\bf r})v_{i}({\bf r'})=E_{i}u_{i}({\bf r})
\label{coupleB1}
\end{eqnarray}
and
\begin{eqnarray}
\hat{L}v_{i} &+&\int d{\bf r'}\,V(|{\bf r}-{\bf r'}|)\Phi^{*}({\bf r})
\Phi({\bf r'})v_{i}({\bf r'})\nonumber \\
&+&\int d{\bf r'}\,V(|{\bf r}-{\bf r'}|)
\Phi^{*}({\bf r})\Phi^{*}({\bf r})u_{i}({\bf r'})=-E_{i}v_{i}({\bf r})\,.
\label{coupleB2}
\end{eqnarray}
Hence with the solution of $u_{i}$ and $v_{i}$ from the above coupled 
equations, the Hamiltonian in Eq. (\ref{newAuxHamil}) can be shown to reduce to
\begin{equation}
\hat{H}_{new}^{s}= -\sum_{i} E_{i}\int\,d{\bf r}|v_{j}({\bf r})|^2
+\sum_{i}E_{i}\alpha_{i}^{\dagger}\alpha_{i}\,,
\label{reducedH}
\end{equation}
which describes a non-interacting gas of quasiparticles of energy $E_{j}$. 
The ground state expectation value of Eq. (\ref{reducedH}) is given by
\begin{equation}
E_{0}^{s}= -\sum_{i} E_{i}\int\,d{\bf r}|v_{j}({\bf r})|^2\,,
\label{reducedE}
\end{equation}
taking into consideration that the ground state $\ket{g_{s}}$ obeys
$\alpha\ket{g_{s}}=0$. Besides, using Eq. (\ref{bogo}), the local 
non-condensate density defined in Eq. (\ref{noncon}) can be written as 
\begin{equation}
\tilde{n}({\bf r'})\equiv \bra{g_{s}}\tilde{\psi}^{\dagger}({\bf r})
\tilde{\psi}({\bf r})\ket{g_{s}}= \sum_{i}|v_{i}({\bf r})|^2 \,.
\label{noncon2}
\end{equation} 
Thus performing a little algebra, the new auxiliary functional which includes 
depletion of the condensate is defined by
\begin{equation}
F_{s}[n({\bf r})]= \int\,d{\bf r} \Phi^{*}({\bf r}) \left( -\frac{\nabla^2}{2m}
-\mu \right)\Phi({\bf r})- \sum_{i}\int d{\bf r} |v_{i}({\bf r})|^2[E_{j}
+V_{s}({\bf r})] \,.
\label{operator}
\end{equation} 
The key element in the above derivation is that one can take into account the 
depletion of the condensate as shown in Eq. (\ref{noncon2}). Using the the 
auxiliary system defined by the Hamiltonian Eq. (\ref{AuxHamil}) and the 
relation  in Eq (\ref{trapRel}), both the condensate density 
$n_{c}\equiv |\Phi({\bf r})|^2$ and also the total density $n({\bf r})$ can be 
computed.

The difficulties in finding the solution for Eq. (\ref{LifePrimo}), 
Eqs (\ref{coupleB1}) and (\ref{coupleB2}) lies in the fact that the 
exact functional dependence of the exchange-correlation potential 
$v_{xc}[n({\bf r})]=\delta F_{xc}[n({\bf r})]/\delta n({\bf r})$ is 
unknown for most of the system of interest and thus one has to resort to 
approximations such as the Local Density Approximation (LDA). 

\subsection{Application of DFT to N vortex problem}
 
We are now in a position to analyse our problem of N vortices treated as 
N Yukawa bosons via the one-to-one mapping of Nelson and Seung \cite{nelson1} 
with the DFT tool summarized in the previous section within our approriate 
scaling. By choosing an auxiliary system of N non-interacting bosons 
(referring to Eq. (\ref{AuxFunc}) and Eq. (\ref{trapRel})), 
Eq. (\ref{LifePrimo}) then can be written more explicitly as satisfying the 
following equation:
\begin{equation}
\left( -\frac{\hbar^2}{2m}\nabla^2+ V_{ext}({\bf r})+ V_{H}({\bf r})
+\frac{\delta F_{xc}[n({\bf r})]}{\delta n({\bf r})}\right)\Phi({\bf r})
=\mu\Phi({\bf r})\,.
\label{KS}
\end{equation}
This equation is a sort of generalization of Gross-Pitaevskii equation for a 
dilute inhomogeneous gas at $T=0$. Since we have chosen a non-interacting 
auxillary system, we expect that at $T=0$ all the atoms are Bose condensed as 
their total density $n({\bf r})$ is approximated by condensate density 
$n_{c}({\bf r})=|\Phi({\bf r})|^2$. Setting $F_{xc}[n({\bf r})]=0$ one 
recovers a closed non-linear Schr{\" o}dinger equation (NLSE).\par

\begin{figure}
\centering{
\epsfig{file=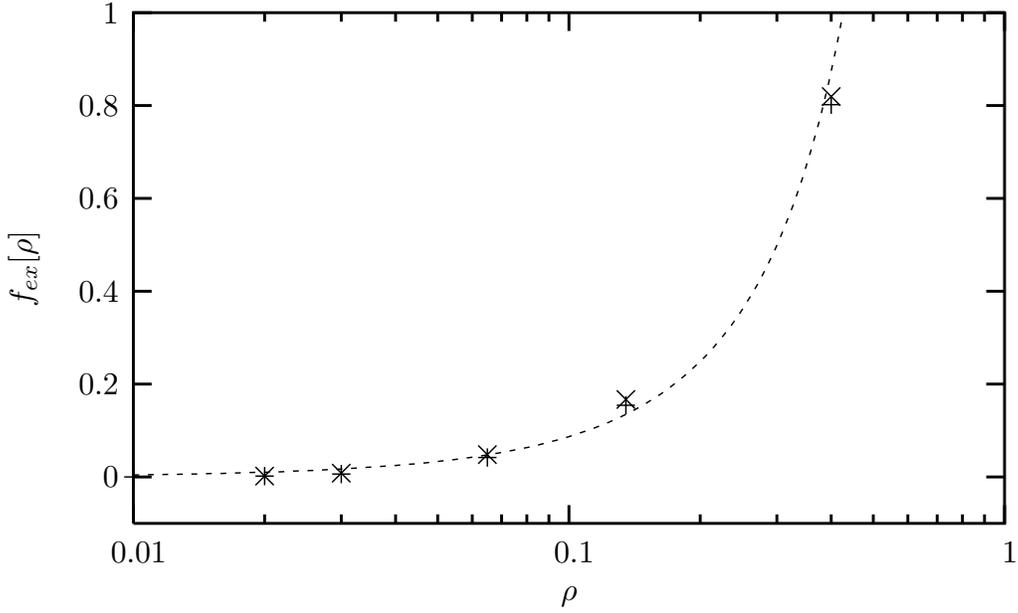,width=1.0\linewidth}}
\caption{Homogeneous excess-correlation energy $f_{ex}(\rho)$ 
versus the dimensionless homogenous density $\rho$ gas of the VMC data of 
Magro and Ceperley \cite{Magro93a} (pluses) and STLS data of Strepparola 
et. al. \cite{strepa} (crosses) compared to the fit function 
Eq. (\ref{gfit}) (dashed-lines).}  
\label{fig_fit}
\end{figure}

The available numerical data  of Magro and Ceperley \cite{Magro93a} and 
Strepparola et. al \cite{strepa} permit one to obtain information on the 
homogeneous excess free energy $ f_{ex}[n]$ rather than the homogeneous 
exchange correlation energy $ f_{xc}[n]$ . Thus it is much convenient to work 
with excess free energy defined as \cite{moroni}
\begin{equation}
F_{ex}[n({\bf r})]= \hat{V}_{H}[n({\bf r})]+F_{xc}[n({\bf r})] \,.
\label{relation}
\end{equation}
In general, the excess correlation functional energy  $F_{ex}[n({\bf r)}]$ 
in a DFT calculation is not known exactly. One can resort to approximations 
such as the Local density approximation (LDA) which reads,
\begin{equation}
 F_{ex}[n({\bf r})]\approx \int\,d{\bf r} F_{ex}^{hom}[n]|_{
n\rightarrow n({\bf r})}\,
\label{LDA}
\end{equation}
where $F_{ex}^{hom}[n]=n f_{ex}^{hom}[n]$. The functional derivatives 
(excess-correlation potential) of the above relation can be 
written as \cite{likos,albus}
 \begin{equation}
V_{ex}({\bf r},n({\bf r}))=\frac{\delta F_{ex}[n({\bf r})]}{\delta  n({\bf r})}
=\frac{\partial (n f_{ex}[n]) }{\partial n}|_{n\rightarrow 
n({\bf r})}\,.
\label{partial}
\end{equation}
Information of the homogeneous excess correlation energy $f_{ex}[n]$ can be 
obtained by subtracting the kinetic energy from the total ground state energy
 of a homogeneous system with N boson. A plot of the DMC data of Magro and 
Ceperley \cite{Magro93a} and the STLS data of Strepparola et. al. \cite{strepa} are depicted along with the following fit
\begin{equation} 
f_{ex}[\rho]= \frac{2\rho}{\ln(1/\rho)}\,,
\label{gfit} 
\end{equation}
in Fig (\ref{fig_fit}). To be consistent with the scaling of Magro and 
Ceperley \cite{Magro93a} I will scale all lengths by harmonic oscillator 
length  $\sigma=\sqrt{\hbar/m\omega_{\perp}}$ and the energy by 
$\epsilon=\hbar\omega_{\perp}/2$. Eq. (\ref{KS}) is solved numerically 
using the excess correlation potentials of Eq.(\ref{partial}) and 
Eq. (\ref{gfit}) self-consistently with the condition that the areal integral 
of the condensate density $|\Phi({\bf r})|^2$ is equal to the 
total number $N$ of bosons.
% However the use of density in the logaritmic term to solve Eq. (\ref{KS}) is not so trivial as thought. To have a stable numerical solution I have subtituted the density parameter in the logaritmic term by $\mu ma^{2}/\hbar^2=4\pi na^{2}/\ln(1/4\pi na^{2})$ obtained from the random walk calculation of Nelson et. al.\cite{nelson1,kolo}. 

\clearpage

\begin{figure}
\centering{
\epsfig{file=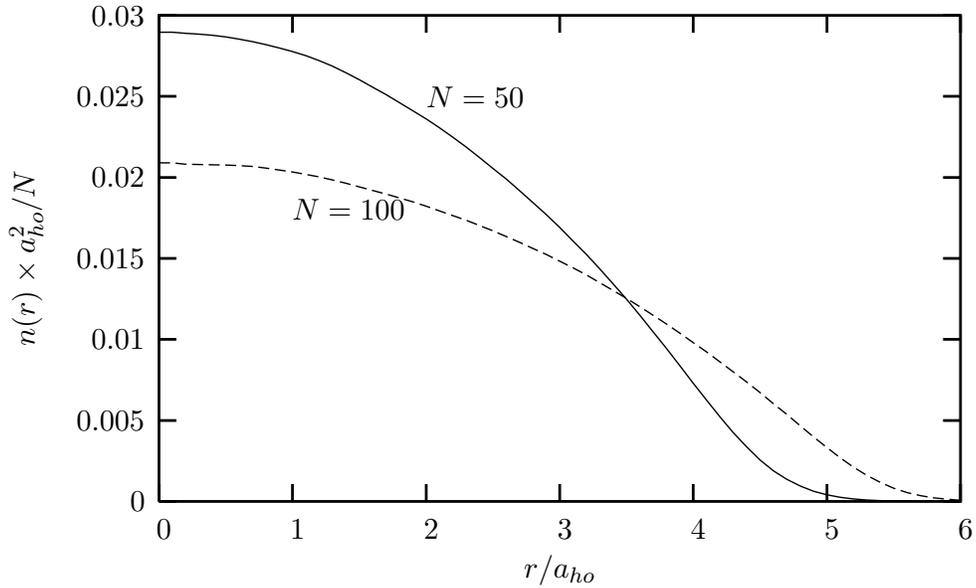,width=1.0\linewidth}}
\caption{Density profiles $n(r)$ for the condensate ( in units $a_{ho}^2/N$, 
with $a_{ho}=\sqrt{\hbar/m\omega_{\perp}}$ ) versus the radial distance $r$ 
(in units $a_{ho}$ ) obtained using Eq.(\ref{KS}) for $N=50$ and $N=100$ 
bosons}
\label{fig_mu}
\end{figure}

%\begin{figure}
%\centering{
%\epsfig{file=fig_1P4.tex.eps,width=1.0\linewidth}}
%\caption{caption as in Fig. (\ref{fig5P3}) for $N=1\times 10^{4}$ bosons.}
%\label{fig1P4}
%\end{figure}

%%%%%%%%%%%%%%%%%%%%%%%%%%%%%%%%%%%%%%%%%%%%%%%%%%%%%%%%%%%%%
%\clearpage{\pagestyle{empty}\cleardoublepage}
\chapter[Summary and Future Direction]{Summary and Future Direction} 
\label{fcs_shuttle_chapter}
\section{Summary and future directions}
\label{future}

In this thesis I have reported on the development on some fundamental 
microscopic aspects of a two-dimensional Bose-Einstein condensate at finite 
temperature. The approach presented here is of mean-field nature which
obeys the spontaneous symmetry breaking. This leads to a nonzero mean value of 
the atomic Bose field operator, which corresponds to the order parameter for 
the BEC phase transistion. The order parameter which describes the mean-field 
condensate is highly correlated with quantum fluctuations. The condensate 
which is well described by the Gross-Pitaevskii equation is coupled with the 
Hartree-Fock model of the thermal cloud. We call this the two-fluid model
description in this thesis.  

Phase fluctuations and scattering properties play a crucial role in the 
studies of 2D Bose gas. One important consequence of lowered dimensionality 
is that the T-matrix for two-body collisions {\it in vacuo} at low momenta and 
energy, which should be used to obtain the collisional coupling parameter to 
the lowest order in the particle density, vanishes in the strictly 2D limit as 
the {\it{s}}-wave scattering length becomes larger than the width of the axial 
trapping \cite{schick,popov}. It is then  necessary to evaluate the scattering 
processes between pairs of Bose particles in by taking into account the 
presence of a condensate and a thermal cloud through a many-body T-matrix 
formalism \cite{stoof93,bijlsma97,Lee}. In fact in this thesis I have 
calculated the many-body T-matrix which includes the effect of thermal 
excitations. I have found that the many-body screening of the interactions due 
to the occupancy of excited states is quite large and rapidly increasing with 
temperature. However such many-body screening has, however, very little effect 
on equillibrium properties of the gas for our choice of system parameters
for the temperature of interest in this work ( $T \le 0.5 \,T_{c}$). This work 
has been published as ``Density profile of a strictly two-dimensional Bose 
gas at finite temperatur'', K. K. Rajagopal, P. Vignolo and M. P. Tosi, 
Physica B {\bf 344},157 (2004).

Measurement of Feshbach resonances provides a means of obtaining 
information about interaction strength (repulsive $g>0$ or attractive $g>0$)
between atoms. The exact location and width are very important in deducing the
interatomic potentials, but to our knowledge they have not been calculated or 
measured experimentally for the strictly 2D case. Series of calculation for 
different species of atomic gasses for the 3D cases can be found in the 
literature \cite{moerdijk95,vogels,tiesinga92,tiesinga93,inouye}.

In this thesis I have explicitly applied the formalism already developed to 
treat Feshbach resonances in 3D boson gases to the case of a 2D Bose gas at 
zero temperature. One of the main results is that an external magnetic field 
could be used to drive the coupling from repulsive to attractive even in a 
pancake geometry as the scattering length becomes larger than the axial width 
of the condensed cloud. For the case of repulsive interactions I have 
evaluated the particle density profiles at varying coupling strength, 
demonstrating the approach to a Thomas-Fermi profile followed by collapse. In 
the opposite regime of attractive coupling I have given an estimate for the 
critical number of bosons that can form a stable condensate. I found that this 
critical number is inversely proportional to the absolute magnitude of the 
coupling, independently of the strength of the in-plane confinement. This work 
has been published as ``Feshbach resonances in a strictly two-dimensional 
atomic Bose gas'', K. K. Rajagopal, P. Vignolo and M. P. Tosi, Physica B {\bf 353}, 59 (2004).  

Also in this thesis I have calculated the particle density, ground state
energy and critical frequency that required to nucleate a single stable vortex 
when a 2D condensate put into rotation at various temperatures. My 
calculations illustrate the interplay between the thermal cloud and the 
structure of the condensate in the regions where the radial kinetic energy of 
the condensate is playing an important role (namely, in the outer parts of the 
condensate and in the vortex core) and validated the use of a simple 
expression of the vortex energy for temperatures up to about 0.5 $T_c$.
This work has been published as `` Temperature dependence of the energy of a 
vortex in a two-dimensional Bose gas '', K. K. Rajagopal, P. Vignolo and 
M. P. Tosi, Physics Letter A {\bf 328}, 500 (2004).

The work presented in thesis has given a broad microscopic description of 
homogeneous or trapped 2D Bose gas. Thus it opens up an interesting range of 
further research possibilities. At the same time, the genarality of the 
approach considered here leaves numerous questions unanswered. One of them
which interest me the  most is the dynamics of many vortices in a 2D 
harmonically trapped Bose gas. In Sec. 3.2 I have developed a density 
functional theory (DFT) technique to calibrate the ground state properties of 
the vortex fluid confined in a two-dimensional harmonic trap within the
Bogoliubov-de Gennes reference system. \par
When a non-interacting auxiliary system is inferred in the DFT model, we 
obtained a Kohn-Sham (KS) type equation. I have calculated  numerically the 
density profile and the ground state energy of the vortex fluid using KS 
equation for $N\leq 137$ bosons. The excess-correlation potential in this 
equation is obtained from a  fit to the liquid phase  VMC data of Magro and 
Ceperley \cite{Magro93a} and Strepparola et.al \cite{strepa} for the 
homogeneous ground state energies. 

The crystallization of Yukawa bosons (vortices) can be studied by making an 
analogy to Wigner crystallization \cite{wigner} in a system of electrons. 
Wigner first predicted that a system of electrons in a uniform potential would 
crystallize at low density within the Hartree-Fock (HF) theory. He proposed 
that at sufficiently low density crystallization greatly reduces the 
interaction energy with only a small increase in the kinetic energy. 
The energy lowering in the crystallization derives from the reduction of the 
interaction energy by spatially separating the electrons. Hartree-Fock studies 
have shown a transition from Fermi liquid to Wigner molecule using small 
number of electrons in a harmonic trap \cite{landman,maninen2}. This 
transition has been a hot topic in the last two years and often associated 
with the breaking of rotational symmetry found in the HF calculation of 
quantum dots. Very recently Romanovsky et. al.\cite{roma} have studied 
crystallization of bosons interacting with Coulomb potentials and contact 
potentials in harmonic trap for small number of particles ($N=6$). They have 
demonstrated a two-step method of breaking the symmetry of strongly repelling 
bosons at the HF level followed by a post HF symmetry restoration incorporating a correlation beyond mean-field. The method describes transition from a BEC 
state to a crystalline phase in which the trap plays acrucial role in 
localizing the bosons. The Hartree-Fock solutions breaks the translational 
invariance of the many-body Hamiltonian and bosons are considered to be pinned 
as in the Wigner crystal. Their method can be used to now is to calculate the 
ground state properties ( density profile and energy ) for ($N\leq 137$) Yukawa
boson in the cystalline phase.

%%%%%%%%%%%%%%%%%%%%%%%%%%%%%%%%%%%%%%%%%%%%%%%%%%%%%%%%%%%%%%
%\clearpage{\pagestyle{empty}\cleardoublepage}
%\chapter[Ultracold atoms in 2D Optical Lattices]{Ultracold atoms in 2D Optical Lattices} 
%\label{fcs_shuttle_chapter}
%\input{Slave.tex}
%%%%%%%%%%%%%%%%%%%%%%%%%%%%%%%%%%%%%%%%%%%%%%%%%%%%%%%%%%%%%%%
%\clearpage{\pagestyle{empty}\cleardoublepage}
%\chapter[Superfluid-Mott Insulator Transition in 2D]{Superfluid-Mott Insulator Transition
%in 2D} 
%\label{fcs_shuttle_chapter}
%\input{qphase.tex}
%%%%%%%%%%%%%%%%%%%%%%%%%%%%%%%%%%%%%%%%%%%%%%%%%%%%%%%%%%%%
%\clearpage{\pagestyle{empty}\cleardoublepage}
%\pagestyle{spec}
%\chapter*{Conclusions: Further developments 
%\markboth{Further developments}{Further developments}} 
%\addcontentsline{toc}{chapter}{Further developments}
%  () \label{proposal}
%  ()input{chapvort.tex}
%%%%%%%%%%%%%%%%%%%%%%%%%%%%%%%%%%%%%%%%%%%%%%%%%%%%%%%%%%%%

\bibliographystyle{unsrt}

\addcontentsline{toc}{chapter}{Bibliography}
%%% serve per aggiungere al sommario la voce Bibliografia

\end{document}